\documentclass[12pt]{iopart}

\usepackage{amsfonts}
\usepackage{iopams}

\newtheorem{proposition}{Proposition}

\begin{document}

\title{A pedagogical presentation of a $C^\star -$algebraic approach to
quantum tomography}
\date{}
\author{A. Ibort$^a$, V.I. Man'ko$^b$, G. Marmo$^{a,c}$, A. Simoni$^c$, F.
Ventriglia$^c$}

\address{$^a$ Departamento de Matem\'{a}ticas, Universidad.
Carlos III de Madrid, Avda. de la Universidad 30, 28911 Legan\'{e}s,
Madrid, Spain }

\address{$^b$ P.N. Lebedev Physical Institute, Leninskii
Prospect 53, Moscow 119991, Russia}

\address{$^c$ Dipartimento di Scienze Fisiche dell'
Universit\`{a} ``Federico II" e Sezione INFN di Napoli,
Complesso Universitario di Monte S. Angelo, via
Cintia, 80126 Naples, Italy}

\eads{\mailto{albertoi@math.uc3m.es}, \mailto{manko@na.infn.it},  \mailto{marmo@na.infn.it},  \mailto{simoni@na.infn.it},
\mailto{ventriglia@na.infn.it}}

\begin{abstract}
It is now well established that quantum tomography provides an alternative
picture of quantum mechanics. It is common to introduce tomographic concepts
starting with the Schrodinger-Dirac picture of quantum mechanics on Hilbert
spaces. In this picture states are a primary concept and observables are
derived from them. On the other hand, the Heisenberg picture,which has
evolved in the $C^\star-$algebraic approach to quantum mechanics, starts
with the algebra of observables and introduce states as a derived concept.
The equivalence between these two pictures amounts essentially, to the
Gelfand-Naimark-Segal construction. In this construction, the abstract $%
C^\star-$algebra is realized as an algebra of operators acting on a
constructed Hilbert space. The representation one defines may be reducible
or irreducible, but in either case it allows to identify an unitary group
associated with the $C^\star-$algebra by means of its invertible elements.
In this picture both states and observables are appropriate functions on the
group, it follows that also quantum tomograms are strictly related with
appropriate functions (positive-type)on the group. In this paper we present,
by means of very simple examples, the tomographic description emerging from
the set of ideas connected with the $C^\star-$algebra picture of quantum
mechanics. In particular, the tomographic probability distributions are
introduced for finite and compact groups and an autonomous criterion to
recognize a given probability distribution as a tomogram of quantum state is
formulated.
\end{abstract}


\maketitle

\section{Introduction}

The problem of quantum state description was the subject of intensive
investigations from the very beginning of quantum mechanics \cite%
{Schroed1926,Landau1927,VonNeumann1927,Wigner1932,Husimi1940,Sudarshan1963,Glauber1963,Glauber1963bis}%
. The association of quantum states with quasi-distributions \cite%
{Wigner1932,Husimi1940,Sudarshan1963,Glauber1963} made the description of
states in quantum mechanics similar to the description of classical particle
states in classical statistical mechanics by means of probability
distributions on phase space. However, the class of quasi-distributions
introduced in quantum mechanics cannot contain all classical distribution
functions because of the uncertainty relation \cite%
{Heisenb1927,Schroed1930,Robertson1930}.

In view of the uncertainty relations there cannot exist a joint probability
distribution function for instance of two random position and momentum since
they cannot be measured simultaneously. It is admissible to have a
probability distribution function of only one of the two conjugate
variables, for instance position.

The problem of discussing the position probability distribution together
with the momentum probability distribution was discussed by Pauli \cite%
{Pauli1935}. Although this problem as formulated by Pauli found negative
solution, it triggered investigations in this direction and it turned out
that one can introduce a family of actual probability distributions of one
random variable (position) called tomographic probability distributions or
simply tomograms, these distributions provide a description of quantum
states in complete analogy with the description of states in classical
statistical mechanics \cite{Mancini96}, see also the recent review \cite%
{Pedatom} and \cite{Ventri1,Ventri2,Ventri3}. It is worthy to note that in
\cite{Ali-Prug,Ali-Prug2,LahtiBush,Stulpebook} the probability approach to
describe the quantum states was discussed but the tomographic version of
such description has appeared only as a result of thinking on experiments on
homodyne detection of quantum photon states \cite{Raymer,Raymer2} based on
optical tomograms whose relation with the Wigner functions was found in \cite%
{Ber-Ber,VogRis}. In these papers the tomograms, being measurable
probability distributions, were considered as a technical tool to measure
the photon quantum states identified with the Wigner functions. So,
tomograms were not considered as primary objects providing an alternative
picture of Quantum Mechanics.

In the papers \cite{Mancini96} and \cite{Ventri1,Ventri2,Ventri3} a new
element in the tomographic approach to quantum mechanics appeared in the
sense that the tomographic distribution itself is identified with the
quantum state. In other words, knowing a quantum tomogram one can obtain all
the quantities of quantum mechanics like the energy spectrum, quantum
transition probabilities, quantum state evolution in the form of an equation
for the probability distribution, etc. Thus the tomogram can be used as
alternative to such primary concept of state as the notion of wave function
or density operator (we also call it a density state).

According to the tomographic approach, for any density state (or wave
function) one constructs the tomogram and \textit{vice versa}, from any
given tomogram one can reconstruct the quantum state density operator $\rho
. $ The density operator $\rho $ has the properties: Hermiticity, i.e. $\rho
^{\dagger }=\rho ,$ trace normalization $\mathrm{Tr}\left[ \rho \right] =1$
and non-negativity, i.e. $\rho \geq 0.$ The tomographic probability
distribution provides the density operator by using the inversion formulae
that are available in explicit form for all kinds of tomograms like the
optical one \cite{D'Ariano}, symplectic tomogram \cite%
{D'Ariano-Mancini,Manko-Tombesi}, spin tomogram \cite{OlgaJETP}, photon
number tomogram \cite{ManciniEPL} and center of mass tomogram \cite{Arkhipov}%
. The problem of measuring continuous position and momentum in connection
with the tomographic description of quantum states was discussed in \cite%
{LahtiKiu} and the discrete spin variables were considered in an anologous
representation in \cite{Weig}.

If we consider from the very beginning the tomograms as conceptual primary
objects associated with quantum states, the question arises for finding
conditions to recognize whether a given probability distribution is a
tomogram of a quantum state. The common answer to this question is that one
has to use the inversion formula to obtain an operator $\tilde{\rho}$ and
then to check if it has all the properties characterizing density operators.
However this answer is unsatisfactory, because it requires to switch from
the tomographic description to other pictures of quantum mechanics.
Analogous problem was considered for Wigner functions and the criterion was
formulated in \cite{Narcowich} on the base of the so called
Kastler-Loupias-Miracle Sole (KLM) conditions \cite{KLM1,KLM2,KLM3}. The
connection of tomograms with Wigner functions could be used \cite%
{VentriPositive}, but again it would be unsatisfactory. One needs autonomous
criteria to answer the question.

In the present paper we provide self-contained conditions for a probability
distribution to be a tomogram of a quantum state, i.e., a quantum tomogram.
We will formulate such properties by using the Naimark method \cite{Naimark}
and the Gelfand-Naimark-Segal (GNS) construction \cite{GNS} to describe
quantum states in terms of vectors of a suitable Hilbert space.

It is worthy to note that the tomographic approach can be formulated in the
framework of a star-product scheme \cite{JPA2002,PatriziaPLA,PatriziaPLA2}.

The strategy of our work is to find the connection between functions (which
are diagonal matrix elements of a unitary representation of a group $G)$ and
quantum tomograms. Any such a function in \cite{Naimark} was shown to have
properties of positivity, recalled in the following. Then, based on this
property and in view of the connection with tomograms, we can establish the
properties characterizing quantum tomograms among other probability
distributions.

The paper is organized as follows. In sections 2 and 3 introductory remarks
on $C^{\star }-$algebras and a simple example are discussed. A concrete case
of $C^{\star }-$algebra, the group algebra, is discussed in section 4 for a
finite group. The fundamental notion of positive-type group function is
recalled in section 5. Section 6 is devoted to introduce the tomographic
descriptions of quantum states based on irreducible representations of a
finite group, \textit{via} a positive group function. Section 6 is the core
of this paper: its definitions and results, which are discussed with extreme
detail in the case of the group of permutations of three points $S_{3}$ in
the long section 7, are straightforwardly extended to the compact groups
like $U(n)$ in section 8, after a \textit{caveat} on the necessity of using
the Gelfand-Zetlin bases \cite{GelfandZetlin,Baird-Biede}. Also, the
tomographic reconstruction formula provided in section 8 is evaluated in
detail for the case of $SU(2).$ The Gelfand-Zetlin bases are discussed with
some care in section 9. The paradigmatic case of $SU(3)$ illustrates the
previous results in section 10. In section 11 the important necessary and
sufficient conditions for a given family of stochastic vectors to be a
tomogram are formulated in terms of a suitable positive-type group function,
both for finite and compact groups. An example based on $S_{3}$ illustrates
the theory. Moreover, the possibility of checking the positivity of a
compact group function \textit{via} the restriction to a finite subgroup is
analyzed. Finally, in section 12 some conclusions and perspectives are drawn.

\section{Introductory remarks on $C^{\star}-$algebras}

It is known that the formulation of quantum mechanics stemming from
Heisenberg picture is given by using a $C^{\star }-$algebra formalism \cite%
{Strocchi}. In this formalism from the very beginning one does not use
neither a Hilbert space nor operators. Instead, it is used an associative
algebra $\mathcal{A}$ with identity $E$ and a $\star -$involution, such that
$(AB)^{\star }=B^{\star }A^{\star },$ plus an appropriate norm $|| \cdot ||$
to introduce a topology. The norm $|| \cdot ||$ satisfies the continuity
requirement for the product $\left\Vert AB\right\Vert \leq \left\Vert
A\right\Vert \left\Vert B\right\Vert $ and the compatibility condition $%
\left\Vert A^{\star }A\right\Vert =\left\Vert A\right\Vert ^{2}.$

The observables of the theory are real (also called self--adjoint) elements:
$A^{\star }=A$. States are normalized positive continuous linear functionals
$\rho $ on this algebra, this is, continuous linear maps such that $\rho
(A^{\star }A)\geq 0$, and $\rho (E)=1,$ (replacing the trace property for
density states). The mean value of an observable $A$ in the state $\rho ,$
say $\langle A\rangle _{\rho },$ is just the real number $\rho (A),$ the
evaluation of $\rho $ on $A.$

Some elements of the algebra $\mathcal{A}$ have an inverse. The elements $U$
for which $U^{\star }=U^{-1}$ are called the unitary elements in the $%
C^{\star }-$algebra, and they form a group $\mathcal{U}$.

Starting from a $C^{\star }-$algebra $\mathcal{A}$, the
Gelfand-Naimark-Segal (GNS) construction provides, given a fiducial state $%
\rho ,$ a Hilbert space carrying a $^\star-$cyclic representation $\Pi$ of
the algebra, $\Pi(A^{\star})=(\Pi(A))^{\dagger}$. In this way one gets
density operators for states and Hermitian operators for observables of the
usual formulation of quantum mechanics.

One of the aim of our work is to introduce the tomographic approach at the
level of the $C^{\star}-$algebra formulation of quantum mechanics and to
relate it with standard formulation by means of the GNS construction.

The idea of a tomographic picture in a $C^{\star }-$algebra is based on the
possibility to represent an observable $A$, at least for group algebras
based on compact groups as it will be discussed in the following, as real
linear combination of projectors, this is in the form
\begin{equation}  \label{decomposition}
A=\sum_{\alpha ,k}\lambda _{k}^{\alpha }P_{k}^{\alpha },
\end{equation}
where $\lambda_k^\alpha$ are real numbers, the observables $P_{k}^{\alpha
}=P_{k}^{\alpha \star }$ are such that $P_{k}^{\alpha }P_{j}^{\beta
}=P_{k}^{\alpha }\delta _{\alpha \beta }\delta _{kj},$ and satisfying $%
\sum_{\alpha ,k}P_{k}^{\alpha }=E$. It follows $\ AP_{k}^{\alpha }=\lambda
_{k}^{\alpha }P_{k}^{\alpha }.$

The same kind of decomposition (\ref{decomposition}) for a g $U$ gives $%
\lambda _{k}^{\alpha }=\exp (i\theta _{k}^{\alpha }),\theta _{k}^{\alpha
}\in \mathbb{R}.$ Now for any state $\rho ,\rho (P_{k}^{\alpha })=\rho
(P_{k}^{\alpha \star }P_{k}^{\alpha })\geq 0,$ so that we may interpret the
formula $\rho (U)=\sum_{\alpha ,k}\exp (i\theta _{k}^{\alpha })\rho
(P_{k}^{\alpha })$ as the evaluation of the state $\rho $ in $U$, providing
the value of each random phase $\theta _{k}^{\alpha }$ with probability $%
W_{k}^{\alpha }(\rho ,U):=\rho (P_{k}^{\alpha })$. In other words, we have
thus defined the tomographic probability $W_{k}^{\alpha }(\rho ,U)$ of
random index $k$ for any given $\alpha $, and write $\langle U\rangle _{\rho
}=:\sum_{\alpha ,k}\exp (i\theta _{k}^{\alpha })W_{k}^{\alpha }(\rho ,U).$

We complete the construction by introducing the notion of the Naimark matrix
$\mathcal{N}_{ij}=\rho (U_{i}^{-1}U_{j}),$ where $i,j$ vary over any finite
set of natural numbers. If it is positive semi-definite, that is $\sum_{i,j}
\mathcal{N}_{ij} \bar{\xi}^i \xi ^j \geq 0$ for all $\xi^i \in \mathbb{C}$,
by definition, $\rho (U)$ is a positive-type function on the group $\mathcal{%
U}$. Finally, particular realizations of $C^{\star}-$algebras as unitary
irreducible representations of different groups provide corresponding
standard definitions of tomography. The use of $C^\star-$algebras
constructed from groups makes possible explicit state reconstruction from
its tomogram.

\section{An introductory example}

In this section we illustrate the notion of $C^{\star}-$algebra by
considering a simple finite dimensional example. Given three orthonormal
vectors in a Hilbert space $\left\vert a_{1}\right\rangle ,\left\vert
a_{2}\right\rangle ,\left\vert a_{3}\right\rangle ,$ let us consider the
linear space $\mathcal{A}_9$ with nine base vectors organized in a table%
\begin{equation}
\left[
\begin{array}{ccc}
\left\vert a_{1}\right\rangle \left\langle a_{1}\right\vert & \left\vert
a_{1}\right\rangle \left\langle a_{2}\right\vert & \left\vert
a_{1}\right\rangle \left\langle a_{3}\right\vert \\
\left\vert a_{2}\right\rangle \left\langle a_{1}\right\vert & \left\vert
a_{2}\right\rangle \left\langle a_{2}\right\vert & \left\vert
a_{2}\right\rangle \left\langle a_{3}\right\vert \\
\left\vert a_{3}\right\rangle \left\langle a_{1}\right\vert & \left\vert
a_{3}\right\rangle \left\langle a_{2}\right\vert & \left\vert
a_{3}\right\rangle \left\langle a_{3}\right\vert%
\end{array}%
\right] =\left[
\begin{array}{ccc}
A_{1} & A_{2} & A_{3} \\
A_{4} & A_{5} & A_{6} \\
A_{7} & A_{8} & A_{9}%
\end{array}%
\right] .
\end{equation}%
We define the table of products for the base vectors corresponding to the
products of projectors
\begin{equation}
\left\vert a_{j}\right\rangle \left\langle a_{m}\right\vert \left\vert
a_{n}\right\rangle \left\langle a_{k}\right\vert =\delta _{m,n}\left\vert
a_{j}\right\rangle \left\langle a_{k}\right\vert .
\end{equation}%
The multiplication of the algebra elements is determined by multiplication
of the basis vectors, whose multiplication table reads:

\begin{equation}
\begin{array}{cc}
&
\begin{array}{ccccccccc}
A_{1} & A_{2} & A_{3} & A_{4} & A_{5} & A_{6} & A_{7} & A_{8} & A_{9}%
\end{array}
\\
\begin{array}{c}
A_{1} \\
A_{2} \\
A_{3} \\
A_{4} \\
A_{5} \\
A_{6} \\
A_{7} \\
A_{8} \\
A_{9}%
\end{array}
& {\left[
\begin{array}{ccccccccc}
A_{1} & A_{2} & A_{3} & 0 & 0 & 0 & 0 & 0 & 0 \\
0 & 0 & 0 & A_{1} & A_{2} & A_{3} & 0 & 0 & 0 \\
0 & 0 & 0 & 0 & 0 & 0 & A_{1} & A_{2} & A_{3} \\
A_{4} & A_{5} & A_{6} & 0 & 0 & 0 & 0 & 0 & 0 \\
0 & 0 & 0 & A_{4} & A_{5} & A_{6} & 0 & 0 & 0 \\
0 & 0 & 0 & 0 & 0 & 0 & A_{4} & A_{5} & A_{6} \\
A_{7} & A_{8} & A_{9} & 0 & 0 & 0 & 0 & 0 & 0 \\
0 & 0 & 0 & A_{7} & A_{8} & A_{9} & 0 & 0 & 0 \\
0 & 0 & 0 & 0 & 0 & 0 & A_{7} & A_{8} & A_{9}%
\end{array}%
\right] }%
\end{array}%
\end{equation}

This multiplication table defines the structure constants of the algebra
according to
\begin{equation}
A_{j}\cdot A_{k}=\gamma _{jk}^{l}A_{l}.
\end{equation}%
These structure constants satisfy the quadratic equations arising from
associativity $\gamma _{jk}^{m}\gamma _{ml}^{r}=\gamma _{kl}^{m}\gamma
_{jm}^{r}$. The association
\begin{equation}
A_{j}\Rightarrow (\gamma_j)_k^l := \gamma _{jk}^{l}
\end{equation}%
provides a realization of the algebra $\mathcal{A}_9$ in terms of $9\times 9$
matrices.

Any element $A$\ of the $C^{\star}-$algebra is defined as a complex linear
combination:%
\begin{equation}
A=\sum_{m=1}^{9}c_{m}A_{m}=\sum_{j,k=1}^{3}c_{jk}\left\vert
a_{j}\right\rangle \left\langle a_{k}\right\vert .  \label{dec}
\end{equation}%
The zero vector is given by $\{c_{m}=0\}$ or equivalently $\{c_{jk}=0\}.$
The $\star -$involution is defined by:%
\begin{equation}
\left( \left\vert a_{j}\right\rangle \left\langle a_{k}\right\vert \right)
^{\star }=\left\vert a_{k}\right\rangle \left\langle a_{j}\right\vert ,
\end{equation}%
what implies:%
\begin{equation}
A_{1}^{\star }=A_{1},\quad A_{2}^{\star }=A_{4}, \quad A_{3}^{\star }=A_{7},
\quad A_{5}^{\star }=A_{5}, \quad A_{6}^{\star }=A_{8}, \quad A_{9}^{\star
}=A_{9}.
\end{equation}%
In general, for a linear combination (\ref{dec}), we define:
\begin{equation}
A^{\star }=\sum_{m=1}^{9}c_{m}^{\ast }A_{m}^{\star
}=\sum_{j,k=1}^{3}c_{jk}^{\ast }\left\vert a_{k}\right\rangle \left\langle
a_{j}\right\vert .
\end{equation}%
In view of the product table above, the unity element $E$ satisfying $%
AE=EA=A,$ for any $A,$ is defined as
\begin{equation}
E=A_{1}+A_{5}+A_{9}.
\end{equation}%
As for the norm we may use the usual operator matrix norm.

The unitary elements $U$ in the algebra $\mathcal{A}_9$, satisfying $%
U^{\star }U=UU^{\star }=E,$ are the elements such that $U^{\star }=U^{-1}.$
The inverse element $A^{-1}$ of $A$ given by Eq.(\ref{dec}) exists if and
only if (\textsl{iff}) $\det (c_{jk})\neq 0.$ Moreover, the element is
unitary \textsl{iff} the representative matrix $(c_{jk})$ is unitary, i.e.:
\begin{equation}
\sum_{j=1}^{3}c_{hj}c_{kj}^{\ast }=\delta _{h,k}.
\end{equation}
So, the unitary group $\mathcal{U}$ is isomorphic to the group $U(3).$

To any state $\rho $ we can associated a vector $A_{\rho }$ in the $C^\star
- $algebra $\mathcal{A}_9$ by means of the formula:
\begin{equation}
A_{\rho }=\sum_{m=1}^{9}\rho \left( A^\star_{m}\right)
A_{m}=\sum_{j,k=1}^{3}\rho \left( \left\vert a_{k}\right\rangle \left\langle
a_{j}\right\vert \right) \left\vert a_{j}\right\rangle \left\langle
a_{k}\right\vert .
\end{equation}%
The vector $A_{\rho }$ is real, $A_{\rho }^{\star }=A_{\rho },$ because
states must be hermitian functionals $\rho(A^\star) = [\rho(A)]^*$.
Moreover, the hermitian matrix $(\rho _{jk})=\rho \left( \left\vert
a_{k}\right\rangle \left\langle a_{j}\right\vert \right) $\ has trace one
because $\rho (E) = \rho (A_1 + A_3 + A_5) = \rho (\sum_{k= 1}^3 | a_k
\rangle \langle a_k | ) = \sum_{k = 1}^3 \rho_{kk}$. The matrix $\rho_{jk}$
is definite positive because $\rho(A^\star A) = \sum_{j,m,k,n = 1}^9
c_{jm}^* c_{kn} \rho(|a_{m}\rangle \langle a_j | |a_k \rangle \langle a_{n}
|) = \sum_{k,m,n = 1}^9 c_{km}^* c_{kn} \rho_{mn}$ for arbitrary $c_{km}$.
The positive definite property of the matrix $\rho_{kj}$ can also be seen by
considering the orbits of the transitive action of the unitary group $%
\mathcal{U}$ through the vectors. Thus if we consider a diagonal state $\rho
^{0},$ i.e., a state such that $\rho(|a_n \rangle \langle a_n |) = \rho_n ^0
\geq 0$, $n = 1,2,3$, and zero otherwise, we have:%
\begin{equation}
A_{\rho ^{0}}=\sum_{n=1}^{3}\rho ^{0}\left( \left\vert a_{n}\right\rangle
\left\langle a_{n}\right\vert \right) \left\vert a_{n}\right\rangle
\left\langle a_{n}\right\vert =\sum_{n=1}^{3}\rho _{n}^{0}\left\vert
a_{n}\right\rangle \left\langle a_{n}\right\vert ,
\end{equation}%
Now, if $U$ is a unitary element of $\mathcal{A}_9$ with representative
matrix $u$, then we will denote by $\rho_U$ the state $\rho_U(A) =
\rho(U^\star A U)$ which give all the state vectors $A_{\rho }$ as:
\begin{eqnarray}
A_{\rho } &=&\sum_{j,h=1}^{3}u_{jh}\left\vert a_{j}\right\rangle
\left\langle a_{h}\right\vert \sum_{n=1}^{3}\rho _{n}^{0}\left\vert
a_{n}\right\rangle \left\langle a_{n}\right\vert
\sum_{l,k=1}^{3}u_{kl}^{\ast }\left\vert a_{l}\right\rangle \left\langle
a_{k}\right\vert  \nonumber \\
&=&\sum_{j,k=1}^{3}(u\rho ^{0}u^{\dagger })_{jk}\left\vert
a_{j}\right\rangle \left\langle a_{k}\right\vert =\sum_{j,k=1}^{3}\rho
_{jk}\left\vert a_{j}\right\rangle \left\langle a_{k}\right\vert .
\end{eqnarray}%
where $(u^{\dagger }\rho u)_{mn}=\rho _{n}^{0}\delta _{m,n}.$

\textit{Vice versa, }in the dual space of the $C^{\star}-$algebra any vector
$A$ has a dual partner: the base partners are:%
\begin{equation}
\left\vert a_{j}\right\rangle \left\langle a_{k}\right\vert \mapsto
\vartheta _{jk}:\vartheta _{jk}(\left\vert a_{m}\right\rangle \left\langle
a_{n}\right\vert )=\delta _{j,m}\delta _{k,n},
\end{equation}%
so that from Eq. (\ref{dec}) we get:
\begin{equation}
\vartheta _{jk}(A)=c_{jk}
\end{equation}%
and for the partner of $A$:%
\begin{equation}
A\mapsto \alpha _{A}=\sum_{j,k=1}^{3}c_{jk}^{\ast }\vartheta _{jk},
\end{equation}%
giving:%
\begin{equation}
\alpha _{A}(A)=\sum_{j,k=1}^{3}c_{jk}^{\ast }c_{jk}=\mathrm{Tr}\left[
c^{\dagger }c\right] =||A||^{2}.
\end{equation}
Please note that one has to rescale the above Hilbert-Schmidt norm giving $%
||E||^2=3,$ in order to have the property $||E||=1.$ Given any state,
positive-type functions on the unitary group $\mathcal{U}$\ are introduced as%
\begin{equation}
\varphi (U)=\left\langle U\right\rangle _{\rho }=\rho (U).
\end{equation}%
They satisfy the positive semidefinite $n\times n$ matrix condition, for any
$n:$
\begin{equation}
\varphi (U_{j}^{\star }U_{k})\geq 0,\quad \forall U_{j},U_{k}\in \mathcal{U}%
,\quad j,k=1,2,\dots ,n,\forall n.
\end{equation}

The positive-type functions $\varphi (U)$ may be expanded in terms of
tomograms $W_k(\rho, U)$ by diagonalizing the unitary matrix representing $U$
:%
\begin{eqnarray}
\varphi (U) &=&\rho \left( \sum_{j,h}u_{jh}\left\vert a_{j}\right\rangle
\left\langle a_{h}\right\vert \right) = \sum_{j,h,k}v_{jk}\exp (i\theta
_{k})v_{hk}^{\ast }\rho \left(\left\vert a_{j}\right\rangle \left\langle
a_{h}\right\vert \right)  \nonumber \\
&=&\sum_{k}\exp (i\theta _{k})\left( \sum_{j,h}v_{hk}^{\ast }\rho
_{hj}v_{jk}\right) =\sum_{k}\exp (i\theta _{k})W_{k}(\rho ,U)
\end{eqnarray}
Note that the tomogram component $W_{k}(\rho ,U)=(v^\dagger \rho
v)_{kk}\geq0 $ is a component of a stochastic vector: $\sum_{k} (v^\dagger
\rho v)_{kk}=1 $.

\textbf{Remark}. One could have started with two vectors $\left\vert
a_{1}\right\rangle ,\left\vert a_{2}\right\rangle .$ Then the unitary group
of the resulting $C^{\star }-$algebra is isomorphic to $U(2).$ In that case
one can embed the permutation group of three elements into $U(2)$, via a
unitary irreducible representation, so that the corresponding positive-type
functions allow for a tomographic reconstruction of the state $\rho $, as
will be discussed in the following.

\section{Group algebra}

Another example of $C^{\star}-$algebra is the so called group algebra, which
is a tool important \textsl{per se} \cite{Naimark}.

Following \cite{Naimark} we review below some properties of a group algebra,
focusing first on finite groups. Given a finite group of order $%
K:G=G_{K}=\{g_{1},g_{2},\dots g_{K}\},$ consider the complex valued
functions on the group $f:G\rightarrow \mathbb{C}.$ The group algebra
consists of formal linear combinations of group elements:%
\begin{equation}
A_{f}=\sum\limits_{j=1}^{K}f\left( g_{j}\right) g_{j} ,
\end{equation}
and will be denoted by $\mathbb{C}[G]$ or $\mathcal{A}_G$. Each element $A
\in \mathcal{A}_G$ is represented by the coefficients $f(g_i)$ of the
combination and \textit{vice versa}. We have a one-to-one correspondence
between elements of the group algebra and complex valued functions on the
group.

If
\begin{equation}
A_{f}=\sum\limits_{j=1}^{K}f\left( g_{j}\right) g_{j},\quad
A_{h}=\sum\limits_{j=1}^{K}h\left( g_{j}\right) g_{j}
\end{equation}
we have $A_{f}+ A_{h} = A_{f+h}.$ Components of a product are obtained from
\begin{eqnarray}
A_{f} \cdot A_{h} &=&\sum\limits_{j,k}f\left( g_{j}\right) h\left(
g_{k}\right) g_{j}g_{k}=\sum\limits_{j,k}f\left( g_{j}g_{k}^{-1}\right)
h\left( g_{k}\right) g_{j}  \nonumber \\
&=&\sum\limits_{j,k}f\left( g_{j}\right) h\left( g_{j}^{-1}g_{k}\right)
g_{k}= A _{f\cdot h},
\end{eqnarray}%
where, on the algebra of group functions, the convolution product
(star-product) is defined as:
\begin{equation}
(f\cdot h)(g_{k})=\sum_{i=1}^{K}f(g_{j})h(g_{j}^{-1}g_{k}).  \label{ConvAlg}
\end{equation}

The conjugate $A^{\ast }$ of $A$ is defined by setting $g^{\ast }=g$ and
\begin{equation}
A^\ast_{f}=\left(\sum\limits_{j=1}^{K}f\left( g_{j}\right) g_{j}
\right)^\ast=\sum\limits_{j=1}^{K}f^\ast \left( g_{j}\right)
g_{j}=A_{f^\ast},
\end{equation}
i.e. $f^{\ast }(g)=\left[ f(g)\right] ^{\ast }.$

We introduce also the transpose $A^{\mathrm{T}}$ of $A$ by $g^{\mathrm{T}%
}=g^{-1}$ and
\begin{equation}
A^{\mathrm{T}}_{f}=\left(\sum\limits_{j=1}^{K}f\left( g_{j}\right) g_{j}
\right)^{\mathrm{T}}=\sum\limits_{j=1}^{K}f \left( g_{j}\right)
g^{-1}_{j}=\sum\limits_{j=1}^{K}f \left( g^{-1}_{j}\right) g_{j}=A_{f^{%
\mathrm{T}}},
\end{equation}
thus $f^{\mathrm{T}}(g)=f\left( g^{-1}\right) .$ For a product $A\cdot B$ we
have $\left( A\cdot B \right) ^{\mathrm{T}}= B^{\mathrm{T}} \cdot A^{\mathrm{%
T}}.$

Hermitian conjugation is now defined as the composition of conjugation and
transposition: $A^{\star }=\left( A^{\mathrm{T}}\right) ^{\ast }$ or $%
g^{\star }=g^{-1}$ and
\begin{equation}
A^{\star}_{f}=\sum\limits_{j=1}^{K}f^\ast\left( g_{j}\right) g^{-1}_{j}
=\sum\limits_{j=1}^{K}f^\ast\left( g^{-1}_{j}\right) g_{j}=A_{f^{\star}},
\end{equation}
i. e., $f^{\star}(g)=f^{\ast}(g^{-1}).$ For a product $A\cdot B $ we have\ $%
\left( A\cdot B \right) ^{\star }= B^{\star }\cdot A^{\star }.$

All above operations are involutions, this is:
\begin{equation}
\left( A^{\ast }\right) ^{\ast }=A , \quad \left( A^{\mathrm{T}}\right) ^{%
\mathrm{T}}= A ,\quad \left( A^{\star}\right) ^{\star } = A.
\end{equation}

We observe that only the $\star-$involution satisfies the condition $\Pi
\left( A^\star \right)=\Pi^\dagger \left( A \right),$ for any unitary
representation $\Pi$ of the algebra.

The trace of an element $A\in \mathcal{A}_{G}$ is defined by $\mathrm{Tr}%
\left[ g\right] =1$ for $g=e$, the group unity, and $\mathrm{Tr}\left[ g%
\right] =0$ otherwise. We have
\begin{equation}
\mathrm{Tr}\left[ A_{f}\right] =f(e).
\end{equation}%
Now we introduce the scalar product $\langle A_{f},A_{h}\rangle $ in the
group algebra $\mathcal{A}_{G}$ by%
\begin{equation}
\langle A_{f},A_{h}\rangle =\mathrm{Tr}\left[ A_{f}^{\star }\cdot A_{h}%
\right] =\sum\limits_{j=1}^{K}f^{\ast }\left( g_{j}\right) h\left(
g_{j}\right) .
\end{equation}%
which agrees with the standard inner product on complex valued functions on $%
G$ considered as vectors on $\mathbb{C}^{K}$. It follows that $\mathrm{Tr}%
\left[ A_{f}^{\star }\cdot A_{h}\right] =\left( \mathrm{Tr}\left[
A_{h}^{\star }\cdot A_{f}\right] \right) ^{\ast }.$ It is worth to note that
$A^{\dagger }$ is the Hermitian conjugate of $A$ with the scalar product we
have just defined.

The associativity of the group $G$ implies the associativity of the group
algebra. The scalar product is preserved by left and right action, $%
g_{k}\mapsto L_{g_{i}}(g_{k})=g_{j}g_{k},$ $g_{j}\mapsto
R_{g_{k}}(g_{i})=g_{j}g_{k},$ and similarly under conjugation $g_{k}\mapsto
C_{g_{j}}(g_{k})=g_{j}g_{k}g_{j}^{-1}$. It is also invariant under
transposition $A\mathbb{\mapsto }A^{\mathrm{T}}$ and multiplication by a
phase $A\mathbb{\mapsto }\exp \left( \mathrm{i}\theta \right) A$. Under the
transformations $A\mathbb{\mapsto }A^{\ast }$ and $A\mathbb{\mapsto }%
A^{\star }$ the scalar product goes into its complex conjugate. Moreover the
transformations $A\mathbb{\mapsto }gAg^{-1}$ and $A\mathbb{\mapsto }A^{\ast }
$ are automorphisms of the group algebra. We should also mention that a
pointwise product is available:%
\begin{equation}
A_{f}\circ A_{h}=\sum\limits_{j}f\left( g_{j}\right) h\left( g_{j}\right)
g_{j}=A_{fh}
\end{equation}%
which is called the Hadamard product.

\subsection{Representations of group algebras}

All irreducible representations of a finite group $G$ of order $K$ are
finite dimensional and equivalent to unitary representations of it. If $%
\{D^{\alpha }\},$ with $\mathrm{dim} D^{\alpha }_{\alpha }$ is an
irreducible unitary representation of $G$, then because of Schur's Lemma, we
get the orthogonality conditions:
\begin{equation}
\sum_{j=1}^{K}\left( D_{rs}^{\alpha }\left( g_{j}\right) \right) ^{\ast
}D_{pq}^{\beta }\left( g_{j}\right) =\frac{K}{n_{\alpha }}\delta _{\alpha
,\beta }\delta _{r,p}\delta _{s,q},  \label{orthog}
\end{equation}
that imply that the matrix elements $\left\{ D_{rs}^{\alpha }\left(
g_{j}\right) \right\} $ of the set of irreducible unitary representations of
$G$ form an orthogonal set on the algebra $\mathcal{A}_G$. Notice that the
subspace of $\mathcal{A}_G$ spanned by the elements $(r,s)$ of the
irreducible representation $D^\alpha$ is invariant under left (or right)
translations, hence they define invariant subspaces of the regular
representation, i.e., the canonical representation of the group $G$ on its
algebra $\mathcal{A}_G$ by left translations. Hence all irreducible
representations are contained in the regular representation, then there is a
finite number of irreducible representations labelled by $\alpha$ and the
matrix elements $\left\{ D_{rs}^{\alpha }\left( g_{j}\right) \right\} $ form
an orthogonal basis in the algebra of group functions:%
\begin{equation}
f\left( g_{j}\right) =\sum_{\alpha }\sum_{r,s=1}^{n_{\alpha }}c_{rs}^{\alpha
}D_{rs}^{\alpha }\left( g_{j}\right) .
\end{equation}%
Moreover the dimensions $n_\alpha$ of the irreducible representations $%
\{D^{\alpha }\},$ satisfy the equation:%
\begin{equation}
\sum_{\alpha }n_{\alpha }^{2}=K.
\end{equation}

One can use a unitary (reducible or irreducible) representation $U\left(
g\right) $ of the group acting on an $N-$dimensional Hilbert space, to
introduce a representation of the group algebra by means of operators on the
same Hilbert space. The operator $\hat{A}_{f}$ corresponding to the group
algebra element $A_{f}$ will be:
\begin{equation}
\hat{A}_{f}=\sum\limits_{j=1}^{K}f\left( g_{j}\right) U\left( g_{j}\right) .
\end{equation}%
In view of
\begin{equation}
U\left( g_{j}^{-1}g_{l}\right) =U\left( g_{j}^{-1}\right) U\left(
g_{l}\right) ,
\end{equation}%
one gets%
\begin{eqnarray}
\hat{A}_{f} \hat{A}_{h} &=&\sum\limits_{j=1}^{K}f\left( g_{j}\right) U\left(
g_{j}\right) \sum\limits_{l=1}^{K}h\left( g_{j}^{-1}g_{l}\right) U\left(
g_{j}^{-1}g_{l}\right) \\
&=&\sum\limits_{j,l=1}^{K}f\left( g_{j}\right) h\left(
g_{j}^{-1}g_{l}\right) U\left( g_{l}\right) =\sum\limits_{l=1}^{K}\left(
f\cdot h\right) \left( g_{l}\right) U\left( g_{l}\right) = \hat{A}_{f\cdot
g} .  \nonumber
\end{eqnarray}

When $U\left( g\right) $ is an irreducible representation $D^{\alpha }(g)$
the orthogonality relations $\left( \ref{orthog}\right) $ may be used to
obtain the inversion formula%
\begin{equation}
f\left( g\right) = \frac{n_{\alpha }}{K} \, \mathrm{Tr}\left[ \hat{A}%
_{f}D^{\alpha \dagger }\left( g\right) \right] .
\end{equation}

This shows that we are in the framework of a star-product scheme, where
quantizer and dequantizer operators are $D\left( g\right) $ and $D^{\dagger
}\left( g\right)$ respectively \cite{JPA2002,PatriziaPLA,PatriziaPLA2}.

\textbf{Remark.} When the group is finite, the group $\mathcal{U}$ of
unitary elements in the group algebra may be readily determined. $\mathcal{U}
$ consists, by definition, of the elements corresponding to group functions $%
f$'s satisfying the relation
\begin{equation}
f^{-1}=f^{\star },  \label{unitary f}
\end{equation}
where we recall that $f^{\star }$ is defined as $f^{\star }(g)=f(g^{-1}).$ Condition (\ref{unitary f}) expresses unitarity at the abstract
level of group algebra.

This implies that eq.(\ref{unitary f}) is equivalent to the condition of
unitarity for the operator $u_{f}=\sum%
\limits_{j=1}^{K}f\left( g_{j}\right) D(g_{j}),$ for any unitary irreducible
representation of the group.

For finite groups the set of such representations is finite and known, so
condition (\ref{unitary f}) gives explicitly $\mathcal{U}.$ We have
\begin{equation}
f\in \mathcal{U} \quad \leftrightarrow \quad u^{\alpha
}_{f}=\sum\limits_{j=1}^{K}f\left( g_{j}\right) D^{\alpha }(g_{j})\in
U(n_{\alpha })\ ,\forall \alpha  \label{unitary f bis}
\end{equation}%
where $D^{\alpha }$ is an irreducible $n_{\alpha }-$dimensional
representation of the finite group, and $U(n_{\alpha })$ the corresponding
unitary group. When $D^{\alpha }$ varies in the set $\{D^{\alpha }\}$ of all
irreducible representations of the finite group we get a set of $%
\sum_{\alpha }n_{\alpha }^{2}=K$ linear inhomogeneous equations in the $K$
variables $f\left( g_{j}\right) $ with known terms $u_{f}^{1},..,u_{f}^{%
\alpha }\in U(n_{1})\times ...\times U(n_{\alpha }).$ The determinant of
this system does not vanish because its rows are made by matrix elements $%
D_{mn}^{\alpha }(g_{j}),$ an orthonormal set of functions on the group. The
linear system has a unique solution $f_{u^{1},..,u^{\alpha }}$ for any given
$g=u^{1},..,u^{\alpha }\in U(n_{1})\times ...\times U(n_{\alpha })$ and
determines an isomorphism between $U(n_{1})\times ...\times U(n_{\alpha })$
and $\mathcal{U}$ :%
\begin{equation}
f_{g}\cdot f_{h}=f_{gh}\ ,\ \forall g,h\in U(n_{1})\times ...\times
U(n_{\alpha }).
\end{equation}

For instance, in the simplest case of the group $Z^{2}$ there are only 2
representations (one dimensional), $\mathcal{U}$ is isomorphic with $%
S^{1}\times S^{1}$ and the isomorphism is given by
\begin{equation}
\left(e^{i\alpha },e^{i\beta }\right)\in S^{1}\times S^{1}\leftrightarrow
f_{\alpha ,\beta }=\left(\frac{e^{i\alpha }+e^{i\beta }}{2},\frac{e^{i\alpha
}-e^{i\beta }}{2}\right).
\end{equation}%
This result can be easily obtained by solving directly eq.(\ref{unitary f}),
which yields%
\begin{equation}
\frac{1}{f^{2}(g_{1})-f^{2}(g_{2})}%
\left[\begin{array}{c}   
f(g_{1}) \\
-f(g_{2})%
\end{array}\right]%
=%
\left[\begin{array}{c}   
f^{\ast }(g_{1}) \\
f^{\ast }(g_{2})%
\end{array}\right]%
,
\end{equation}%
or the equivalent linear system (\ref{unitary f bis}) which reads:%
\begin{equation}
\left[\begin{array}{c}   
f(g_{1})+f(g_{2}) \\
f(g_{1})-f(g_{2})%
\end{array}\right]%
=%
\left[\begin{array}{c}   
e^{i\alpha } \\
e^{i\beta }%
\end{array}\right]%
.
\end{equation}

\section{Positive-type group functions}

To deal with states and tomograms we need the definition of positive-type
group functions, we recall the definition. Given any group $G,$ a group
function $\varphi \left( g\right) $ is of positive-type if the corresponding
matrix%
\begin{equation}
N_{jk}\left( \varphi \right) :=\varphi \left( g_{j}^{-1}g_{k}\right) ,\quad
g_{j},g_{k}\in \left\{ g_{1},g_{2},\dots ,g_{n}\right\} \subseteq G
\end{equation}%
is positive semidefinite, for any $n-$tuple $\left\{ g_{1},g_{2},\dots
,g_{n}\right\} $\ of elements of $G,$ and for any $n\in \mathbb{N}.$ We may
call $N_{jk}\left( \varphi \right) $ the Naimark matrix of $\varphi .$

For any unitary representation of the group, $U\left( g\right) ,$ it is
possible to define a positive-type group function by means of a pure state
corresponding to the vector $\xi :$%
\begin{equation}
\varphi _{\xi }^{U}\left( g\right) :=\left( \xi ,U\left( g\right) \xi
\right) =\mathrm{Tr}\left[ \rho _{\xi }U\left( g\right) \right] .
\label{pospure}
\end{equation}%
Here, $\rho _{\xi }$ is the density state corresponding to $\xi .$ In fact,
the quadratic form
\begin{eqnarray}
\sum\limits_{j,k=1}^{n}\lambda _{j}^{\ast }\lambda _{k}\mathrm{Tr}\left[
\rho _{\xi }U\left( g_{j}^{-1}g_{k}\right) \right] &=&\mathrm{Tr}\left[ \rho
_{\xi }\sum\limits_{j=1}^{n}\lambda _{j}^{\ast }U^{\dagger }\left(
g_{j}\right) \sum\limits_{k=1}^{n}\lambda _{k}U\left( g_{k}\right) \right] \\
&=&\mathrm{Tr}\left[ \rho _{\xi }V^{\dagger }V\right] \geq 0,  \nonumber
\end{eqnarray}%
where the $\lambda $'s\ are arbitrary complex numbers, is positive
semidefinite.

The above form can be generalized by using any density state $\rho $ instead
of a pure one $\rho _{\xi }$, and this will be very useful in the
tomographic framework.

It should be stressed here that the form of Eq.$\left( \ref{pospure}\right) $
is canonical. Because of Naimark's representation theorem \cite{Naimark},
for any positive-type group function $\varphi \left( g\right) $\ there exist
a Hilbert space, a unitary representation of the group $U\left( g\right) $\
and a cyclic vector $\xi $ such that
\begin{equation}
\varphi \left( g\right) =\left( \xi ,U\left( g\right) \xi \right) .
\end{equation}%
We recall that a vector $\xi $ is called cyclic if the set $\left\{ U\left(
g\right) \xi \mid g \in G \right\} $ spans the Hilbert space.

Notably, for a finite group, the positivity of a group function may be
checked considering only one Naimark matrix, constructed with all the
elements of the group. We have the following

\begin{proposition}
A group function $\psi $\ defined on a finite group $G_{K}=\left\{
g_{1},g_{2},\ldots ,g_{K}\right\} $\ of order $K$ is of positive type
\textsl{iff} the $K\times K-$matrix:%
\begin{equation}
N\left( \psi \right) _{ij}=\psi (g_{i}^{-1}g_{j}),\quad i,j=1,\dots ,K
\end{equation}
is positive semidefinite.
\end{proposition}

{\noindent\emph{Proof:}} In fact, consider the Naimark matrix $N\left( \psi
\right) _{ij}$ of order $K+1$ obtained by adding a repeated element $%
g_{K+1}=g_{h}$ to $\left\{ g_{1},g_{2},\ldots ,g_{K}\right\} .$ Then, $%
N\left( \psi \right) _{ij}$ has two equal rows. So, $\det N\left( \psi
\right) _{ij}=0.$ By induction, $\det N\left( \psi \right) _{ij}=0$ for $%
g_{i},g_{j}\in \left\{ g_{1},\ldots ,g_{K},\ldots ,g_{K+p}\right\} ,$ $%
\forall p,$ and the proposition is proven.

\section{Finite groups and tomography}

We introduce a tomography for any density state $\rho $ by means of a
positive type function on $G,$ defined as:
\begin{equation}
\varphi _{\rho }^{D}(g):=\mathrm{Tr}\left[ \rho D(g)\right]
\end{equation}%
where $\{D(g)\}$ is a unitary representation of $G$ on the Hilbert space
where the density state $\rho$ is defined.

In this section, we consider again a finite group of order $%
K:G=G_{K}=\{g_{1},g_{2},\dots g_{K}\}.$ Then we may suppose the
representation $D$ to be $n-$dimensional ($n^{2}<K $ if $D$ is irreducible.)
For any group element $g$ the corresponding representative matrix $D(g)$ can
be put in the form of a diagonal unitary matrix $d_{g}$ by means of a
unitary matrix $V_{g}:$
\begin{equation}
D(g)=V_{g}d_{g}V_{g}^{\dagger },\quad d_{g}=\mathrm{diag}\left[ \mathrm{e}^{%
\mathrm{i}\theta _{1}(g)},\dots ,\mathrm{e}^{{\mathrm{i}}\theta _{n}(g)}\right]
.
\end{equation}%
We observe that, in general, neither $d_{g}$ nor $V_{g}$ separately are
group representations. Moreover, $V_{g}$ is not uniquely determined for, if $%
C_{r},C_{l}$ are unitary matrices commuting with $d_{g}$ and $\rho $
respectively, we have:
\begin{eqnarray}
\varphi _{\rho }^{D}(g)&:= &\mathrm{Tr}\left[ \rho V_{g}d_{g}V_{g}^{\dagger }%
\right]  \nonumber \\
&=&\mathrm{Tr}\left[ C_{l}^{\dagger }\rho C_{l}V_{g}C_{r}d_{g}C_{r}^{\dagger
}V_{g}^{\dagger }\right]  \nonumber \\
&=&\mathrm{Tr}\left[ (C_{l}V_{g}C_{r})^{\dagger }\rho (C_{l}V_{g}C_{r})d_{g}%
\right] ,
\end{eqnarray}%
so that this ambiguity dos not affect the associate function, and we may
write unambiguously:
\begin{eqnarray}
\varphi _{\rho }^{D}(g) &=&\mathrm{Tr}\left[ d_{g}(V_{g}^{\dagger }\rho
V_{g})\right]  \nonumber \\
&=&\sum_{m=1}^{n}\mathrm{e}^{{\mathrm{i}}\theta _{m}(g)}(V_{g}^{\dagger }\rho
V_{g})_{mm}  \nonumber \\
&=: &\sum_{m=1}^{n}\mathrm{e}^{{\mathrm{i}}\theta _{m}(g)}W_{m}(g,\rho ).
\end{eqnarray}

In the last equation, we have introduced the components
\begin{equation}
W_{m}(g,\rho ):=(V_{g}^{\dagger }\rho V_{g})_{mm}\quad (m=1,\dots ,n)
\end{equation}%
of the vector $\mathbf{W}(g,\rho )$ defining the tomogram of $\rho $ in the
chosen representation of $G_{K}$. We note that, as $V_{g}^{\dagger }\rho
V_{g}$ is again a density state, the tomogram is by definition a stochastic
vector, i.e.:
\begin{eqnarray}
\sum_{m=1}^{n}W_{m}(g,\rho ) &=&\sum_{m=1}^{n}(V_{g}^{\dagger }\rho
V_{g})_{mm}=\mathrm{Tr}[\rho ]=1, \\
W_{m}(g,\rho ) &\geq &0\quad (m=1,\dots ,n),\forall g\in G_{K}.
\end{eqnarray}

The knowledge of the tomograms $\left\{ \mathbf{W}(g_{j},\rho )\right\}
_{j=1}^{K}$ allows for reconstructing the density state. In fact, as the
diagonal matrices $d_{g}$'s depend only on the representation $D$ and are
supposed to be known, the function $\varphi _{\rho }^{D}$ is readily
obtained as:
\begin{equation}
\varphi _{\rho }^{D}(g_{j})=\sum_{m=1}^{n}\mathrm{e}^{\mathrm{i}\theta
_{m}(g_{j})}W_{m}(g_{j},\rho ).
\end{equation}%
Then the state is given by the reconstruction formula:
\begin{equation}
\frac{n}{K}\sum_{j=1}^{K}(\varphi _{\rho }^{D}(g_{j}))^{\ast }D(g_{j})=\rho ,
\label{rec}
\end{equation}%
which is based on the orthogonality relations of the matrix elements of $%
D(g):$
\begin{eqnarray}
&&\frac{n}{K}\sum_{j=1}^{K}(\varphi _{\rho }^{D}(g_{j}))^{\ast
}D_{rs}(g_{j})=\frac{n}{K}\sum_{j=1}^{K}\mathrm{Tr}\left[ \rho ^{\ast
}D^{\ast }(g_{j})\right] D_{rs}(g_{j})  \nonumber \\
&=&\sum_{q,m=1}^{n}\rho _{qm}^{\ast }\frac{n}{K}\sum_{j=1}^{K}D_{mq}^{\ast
}(g_{j})D_{rs}(g_{j})=\sum_{q,m=1}^{n}\rho _{qm}^{\ast }\delta _{m,r}\delta
_{q,s}=\rho _{sr}^{\ast }=\rho _{rs}.
\end{eqnarray}

Now, suppose that $\varphi $ is any positive type function on $G_{K}.$ We
recall that, by Naimark's theorem, there exist a Hilbert space acted upon by
a unitary representation $U$ of $G_{K}$ and a cyclic vector $\xi $ such
that:
\begin{equation}
\varphi (g_{j})=(\xi ,U(g_{j})\xi )=\mathrm{Tr}\left[ \rho _{\xi }U(g_{j})%
\right] .
\end{equation}

In general the above representation $U$ results reducible as a direct sum of
all the irreducible representations $\{D^{\alpha }\},\mathrm{dim}D^{\alpha
}_{\alpha },$ $(\sum_{\alpha }n_{\alpha }^{2}=K),$ of the group, each block $%
D^{\alpha }$ with multiplicity $m_{\alpha }:$%
\begin{equation}
U=\bigoplus_{\alpha }\bigoplus_{s=1}^{m_{\alpha }}D_{s}^{\alpha }.
\end{equation}%
Out of the matrix representing $\rho _{\xi }$ one can extract the same
blocks of the reduction of $U,$ to construct a new matrix $\tilde{\rho},$
with the remaining entries zero. Moreover, $\tilde{\rho}$ is still a state,
as the determinants of its blocks are principal minors of $\rho _{\xi }.$
They are nonzero because $\rho _{\xi }$ is cyclic. Then, by construction,
the function $\mathrm{Tr}\left[ \tilde{\rho}U(g_{j})\right] $ coincides with
the above function $\varphi (g_{j}),$i,e.:%
\begin{equation}
\varphi (g_{j})=\mathrm{Tr}\left[ \rho _{\xi }U(g_{j})\right] =\mathrm{Tr}%
\left[ \tilde{\rho}U(g_{j})\right] .
\end{equation}%
Now we sum together the blocks $\rho _{s}^{\alpha }$ of $\tilde{\rho}$
associated to the same $D^{\alpha }:$%
\begin{equation}
\tilde{\rho}^{\alpha }=\sum_{s=1}^{m_{\alpha }}\rho _{s}^{\alpha }
\end{equation}%
and finally we can write%
\begin{equation}
\varphi (g_{j})=\mathrm{Tr}\left[ \tilde{\rho}U(g_{j})\right] =\sum_{\alpha }%
\mathrm{Tr}\left[ \tilde{\rho}^{\alpha }D^{\alpha }(g_{j})\right] .
\end{equation}

The function $\varphi $ is normalized, i.e., on the identity element $e$ of
the group,$\ \varphi (e)=1.$ Then $\tilde{\rho}^{\alpha }$ can be written as
$\tilde{\rho}^{\alpha }=\gamma ^{\alpha }\rho ^{\alpha },$ where $0\leq
\gamma ^{\alpha }\leq 1,\sum_{\alpha }\gamma ^{\alpha }=1$ and $\rho
^{\alpha }$ is a density state.

So, we have proven:

\begin{proposition}
\label{convex_decomp} Any positive-type function $\varphi $ on $G_{K}$ can
be decomposed as a convex sum of the positive-type functions $\varphi
^{\alpha }$'s related tomographically to the irreducible representations $%
D^{\alpha }$'s of the group:%
\begin{equation}
\varphi (g_{j})=\sum_{\alpha }\gamma ^{\alpha }\varphi ^{\alpha
}(g_{j}),\quad \varphi ^{\alpha }(g_{j})=\mathrm{Tr}\left[ \rho ^{\alpha
}D^{\alpha }(g_{j})\right] .
\end{equation}
\end{proposition}

We remark that any $\varphi ^{\alpha }$ can be written, again using the
Naimark theorem, in terms of a pure cyclic state and a representation $%
U^{\alpha }$ as%
\begin{equation}
\varphi ^{\alpha }(g_{j})=(\xi ^{\alpha },U^{\alpha }(g_{j})\xi ^{\alpha })=%
\mathrm{Tr}\left[ \rho _{\xi ^{\alpha }}U^{\alpha }(g_{j})\right] .
\end{equation}

So, the question arises to relate $U^{\alpha }$ to $D^{\alpha }$ and $\rho
_{\xi ^{\alpha }}$ to $\rho ^{\alpha }.$ Dropping the label $\alpha $, we
can state the following:

\begin{proposition}
\label{reducible} If the density state $\rho $ is of rank $r,$ the\ above
representation $U$ results reducible as a direct sum of $r$ blocks, each one
unitarily equivalent to the irreducible representation $D.$ Then, after a
possible rearrangement, $U=\bigoplus_{s=1}^{r}D_{s}.$ The state $\rho $ can
be used to obtain a pure state $\rho _{\xi },$ cyclic for $U.$
\end{proposition}

The proof amounts to the GNS construction. By using the harmonic expansion
of the group functions in the basis of the matrix elements $D_{qp}^{\beta
}(g_{j})$ of all the irreducible representations $\{D^{\beta }\},\mathrm{dim}%
D^{\beta }_{\beta },$ of the group, and bearing in mind that the dimension
of the given $D$ is $n,$ we may write:
\begin{equation}
\varphi (g_{j})=\sqrt{\frac{n}{K}}\sum_{q,p=1}^{n}\varphi _{qp}D_{qp}(g_{j}),
\end{equation}%
where
\begin{equation}
\varphi _{qp}=\sqrt{\frac{n_{\alpha }}{K}}\sum_{j=1}^{K}\varphi
(g_{j})D_{qp}^{\ast }(g_{j}).
\end{equation}%
Then, the convolution product on the algebra of group functions $\left( \ref%
{ConvAlg}\right) $\ for $X,Y$ reads
\begin{equation}
(X\cdot Y)(g_{j})=\sum_{i=1}^{K}X(g_{i})Y(g_{i}^{-1}g_{j}))
\end{equation}%
and may be expanded as
\begin{eqnarray}
(X\cdot Y)(g_{j}) & = &\sum_{i=1}^{K}\sum_{\alpha }\sqrt{\frac{n_{\alpha }}{K%
}}\sum_{q,p=1}^{n_{\alpha }}X_{qp}^{\alpha }D_{qp}^{\alpha }(g_{i}) \times
\nonumber \\
& & \times \sum_{\beta }\sqrt{\frac{n_{\beta }}{K}}\sum_{m,s=1}^{n_{\beta
}}Y_{ms}^{\beta }D_{ms}^{\beta }(g_{i}^{-1}g_{j})
\end{eqnarray}%
From
\begin{equation}
D_{ms}^{\beta }(g_{i}^{-1}g_{j})=\sum_{r=1}^{n_{\beta }}D_{mr}^{\beta
}(g_{i}^{-1})D_{rs}^{\beta }(g_{j})=\sum_{r=1}^{n_{\beta }}(D_{rm}^{\beta
}(g_{i}))^{\ast }D_{rs}^{\beta }(g_{j})
\end{equation}%
and the orthogonality relations Eq. $\left( \ref{orthog}\right) $, the
convolution product may be written as
\begin{equation}
(X\cdot Y)(g_{j})=\sum_{\alpha }\sum_{q,p,s=1}^{n_{\alpha }}X_{qp}^{\alpha
}Y_{ps}^{\alpha }D_{qs}^{\alpha }(g_{j})=\sum_{\alpha
}\sum_{q,s=1}^{n_{\alpha }}(XY)_{qs}^{\alpha }D_{qs}^{\alpha }(g_{j}).
\end{equation}%
By introducing the function
\begin{equation}
X^{\dagger }(g):=X^{\ast }(g^{-1})=\sum_{\alpha }\sqrt{\frac{n_{\alpha }}{K}}%
\sum_{q,p=1}^{n_{\alpha }}(X_{pq}^{\alpha })^{\ast }D_{qp}^{\alpha }(g),
\end{equation}%
we define a seminorm
\begin{eqnarray}
&&F(X^{\dagger }\cdot X)=\sum_{j=1}^{K}(X^{\star }\cdot X)(g_{j})(\varphi
(g_{j}))^{\ast } \\
\ &=&\sum_{j=1}^{K}\sum_{\alpha }\sum_{q,p,s=1}^{n_{\alpha }}(X_{pq}^{\alpha
})^{\ast }X_{ps}^{\alpha }D_{qs}^{\alpha }(g_{j})\sum_{m,r=1}^{n_{{\alpha }%
_{0}}}\varphi _{mr}^{\ast }(D_{mr}(g_{j}))^{\ast }  \nonumber \\
&=&\frac{K}{n}\sum_{q,p,s=1}^{n}X_{pq}^{\ast }X_{ps}\varphi _{qs}^{\ast }.
\nonumber
\end{eqnarray}%
Without any loss of generality, we may suppose the density state is
diagonal: $\rho =\mathrm{diag}(\rho _{1},\rho _{2},\dots ,\rho _{n}).$ For,
upon diagonalization,
\begin{equation}
\varphi (g):=\mathrm{Tr}\left[ \rho D(g)\right] =\mathrm{Tr}\left[ \mathrm{%
diag}(\rho _{1},\rho _{2},\dots ,\rho _{n})V^{\dagger }D(g)V\right]
\end{equation}%
and we could choose in the previous discussion $V^{\dagger }D(g)V$ instead
of $D(g)$\ from the very beginning. Then
\begin{equation}
\varphi _{qs}^{\ast }=\rho _{q}\delta _{q,s}
\end{equation}%
and the seminorm reads:
\begin{equation}
F(X^{\dagger }\cdot X)=\frac{K}{n}\sum_{q=1}^{n}\left(
\sum_{p=1}^{n}|X_{pq}|^{2}\right) \rho _{q}=\frac{K}{n}\sum_{q=1}^{n}\Vert
\mathbf{X}_{q}\Vert ^{2}\rho _{q}.  \label{seminorm}
\end{equation}%
where the vector $\mathbf{X}_{q}$ is the $q-$th column of the matrix of
coefficients of $D(g)$ in the harmonic expansion of $(X^{\dagger }\cdot X).$

Now, suppose the density state $\rho $ has rank $r,$ with non-zero entries
\begin{equation}
\{\rho _{s_{1}},\rho _{s_{2}},\dots ,\rho _{s_{r}}\}.
\end{equation}%
Then, in view of eq.(\ref{seminorm}), the seminorm kernel $\mathcal{F}%
_{0}=\{X:F(X^{\dagger }\cdot X)=0\}$ is given by the functions $X$ such that
the columns
\begin{equation*}
\{\mathbf{X}_{s_{1}},\mathbf{X}_{s_{2}},\dots ,\mathbf{X}_{s_{r}}\}
\end{equation*}%
of the representative matrix $(X_{pq})$ vanish. So, $F$ is a norm on the
quotient $\mathcal{F}/\mathcal{F}_{0}$ of the algebra of group functions
with respect to the kernel. Equivalence classes are labelled by the entries
of the columns $\{\mathbf{X}_{s_{1}},\mathbf{X}_{s_{2}},\dots ,\mathbf{X}%
_{s_{r}}\}.$ A class representative can be chosen with vanishing expansion
coefficients but those of the above columns of the matrix $(X_{pq})$, we
denote it as $X_{\{\mathbf{X}_{s_{1}},\mathbf{X}_{s_{2}},\dots ,\mathbf{X}%
_{s_{r}}\}}.$ In other words, we have:
\begin{equation*}
X_{\{\mathbf{X}_{s_{1}},\mathbf{X}_{s_{2}},\dots ,\mathbf{X}_{s_{r}}\}}(g)=%
\sqrt{\frac{n}{K}}\sum_{p=1}^{n}\sum_{q=1}^{r}X_{ps_{q}}D_{ps_{q}}(g)
\end{equation*}%
The $r$ columns labeling the classes determine a Hilbert space of dimension $%
rn$, and a corresponding group representation $U^{\ast }$ may be defined as
\begin{eqnarray}
\left( U^{\ast }(h)X_{\{\mathbf{X}_{s_{1}},\mathbf{X}_{s_{2}},\dots ,\mathbf{%
X}_{s_{r}}\}}\right) (g) & := & X_{\{\mathbf{X}_{s_{1}},\mathbf{X}%
_{s_{2}},\dots ,\mathbf{X}_{s_{r}}\}}(h^{-1}g)  \nonumber \\
& = & \sqrt{\frac{n}{K}}\sum_{m=1}^{n}\sum_{q=1}^{r}\left(
\sum_{p=1}^{n}D_{mp}^{\ast }(h)X_{ps_{q}}\right) D_{ms_{q}}(g)  \nonumber \\
& = & X_{\{D^{\ast }(h)\mathbf{X}_{s_{1}},D^{\ast }(h)\mathbf{X}%
_{s_{2}},\dots ,D^{\ast }(h)\mathbf{X}_{s_{r}}\}}(g).
\end{eqnarray}%
In other words, we have $U^{\ast }=\oplus _{s=1}^{r}D_{s}^{\ast },$ or $%
U=\oplus _{s=1}^{r}D_{s},$ and the sum has $r$ terms.

We can use generalized orthogonality relations, to get%
\begin{equation}
\frac{n}{K}\sum_{j=1}^{K}(\varphi (g_{j}))^{\ast
}U(g_{j})=\bigoplus_{s=1}^{r}\rho _{s},\quad \rho _{s}=\rho \quad \forall s
\end{equation}%
where the sum, which has $r$ terms equal to $\rho ,$ is not a density state
any further.

Now, we construct a $rn-$dimensional column vector state $\xi =\{\xi
_{m}\}_{m}$\ by using the nonzero rows of $\rho :$%
\begin{equation}
\xi _{m}=\sum_{j=1}^{r}\sqrt{\rho _{s_{j}}}\delta _{m,n(j-1)+s_{j}},\quad
m=1,2,\dots ,rn
\end{equation}%
which defines a pure state $\rho _{\xi },$ cyclic for $U$ and such that $%
(\xi ,U(g)\xi )=\mathrm{Tr}\left[ \rho D(g)\right] .$

This completes the proof.

\section{The example of $S_{3}$: the permutation group of three elements}

We examine now the group $S_{3}$ of permutation of three elements, which is
isomorphic to the group of symmetries of a triangle, to show how the
considerations of the previous sections appear in a concrete example.

$S_{3}$ has six elements $\left\{ g_{k}:k=1,..,6\right\} $ with a law of
multiplication encoded in the following table

\begin{equation}
R=\left[%
\begin{array}{cccccc}
1 & 2 & 3 & 4 & 5 & 6 \\
2 & 3 & 1 & 5 & 6 & 4 \\
3 & 1 & 2 & 6 & 4 & 5 \\
4 & 6 & 5 & 1 & 3 & 2 \\
5 & 4 & 6 & 2 & 1 & 3 \\
6 & 5 & 4 & 3 & 2 & 1%
\end{array}%
\right]  \label{mult. table}
\end{equation}%
from which one can obtain the group law via
\begin{equation}
g_{i}g_{k}=g_{_{R_{i,k}}}.
\end{equation}%
For example, the table gives $2\cdot 3=1.$ The inverse elements are given by
\begin{equation}
g_{1}^{-1}=g_{1},g_{2}^{-1}=g_{3},g_{3}^{-1}=g_{2},g_{4}^{-1}=g_{4},g_{5}^{-1}=g_{5},g_{6}^{-1}=g_{6}.
\label{inverstable}
\end{equation}%
For example, from this rule we get $2^{-1}=3$. From tables \ref{mult. table}
and \ref{inverstable} one get the table for $g_{i}^{-1}g_{k}$ which reads:
\begin{equation}
L=\left[%
\begin{array}{cccccc}
1 & 2 & 3 & 4 & 5 & 6 \\
3 & 1 & 2 & 6 & 4 & 5 \\
2 & 3 & 1 & 5 & 6 & 4 \\
4 & 6 & 5 & 1 & 3 & 2 \\
5 & 4 & 6 & 2 & 1 & 3 \\
6 & 5 & 4 & 3 & 2 & 1%
\end{array}%
\right] .  \label{NaimTable}
\end{equation}

The space of group functions $f$ on $S_{3}$ is isomorphic to $\mathbb{C}%
^{6}:\ f_{k}:=f(g_{k})\in \mathbb{C};\quad k=1,...,6.$ The Naimark matrix of
$f$ is the$\ 6\times 6-$matrix with entries $f(g_{k}^{-1}g_{m})=f\left(
g_{L_{km}}\right) $ obtained by computing $f$ in the points labelled by $L.$

The left regular representation $D^{L}$ of the group acting on functions $f$
is defined as
\begin{equation}
\left( D^{L}(g_{k})f\right)
(g_{m})=f_{g_{k}}(g_{m})=f(g_{k}^{-1}g_{m})=f\left( g_{L_{km}}\right)
\end{equation}%
therefore in the $k-$th row of the matrix $Lf$ one finds the six values of $%
D^{L}(g_{k})f;$ the left regular representation is made by the following six
$6\times 6-$matrices $D^{L}(g_{k})_{mn}=\delta _{m,L_{kn}}.$ In analogous
way, by using the transpose of $R$ instead of $L,$ one gets the right
action. The characters of the left regular representation are easily
computed to be $\chi ^{L}=(6,0,0,0,0,0).$

$S_{3}$ has three unitary irreducible representations, $D^{0}:\{1,1,1,1,1,1%
\},$ with character $\chi ^{0}:\{1,1,1,1,1,1\},\ D^{1}=\left\{
1,1,1,-1,-1,-1\right\} ,$ with character $\chi ^{1}=\{1,1,1,-1,-1,-1\},$ and
$D^{2}:$
\begin{equation*}
\hskip -1cm \left\{ \left(
\begin{array}{cc}
1 & 0 \\
0 & 1%
\end{array}%
\right) ,\left(
\begin{array}{cc}
\lambda & 0 \\
0 & \overline{\lambda }%
\end{array}%
\right) ,\left(
\begin{array}{cc}
\lambda ^{2} & 0 \\
0 & \overline{\lambda }^{2}%
\end{array}%
\right) ,\left(
\begin{array}{cc}
0 & 1 \\
1 & 0%
\end{array}%
\right) ,\left(
\begin{array}{cc}
0 & \lambda \\
\overline{\lambda } & 0%
\end{array}%
\right) ,\left(
\begin{array}{cc}
0 & \lambda ^{2} \\
\overline{\lambda }^{2} & 0%
\end{array}%
\right) \right\}
\end{equation*}%
where%
\begin{equation*}
\lambda =\mathrm{e}^{\mathrm{i}\frac{2\pi }{3}}=-\frac{1}{2}+\mathrm{i}\frac{%
\sqrt{3}}{2},
\end{equation*}%
with character
\begin{equation*}
\chi ^{2}=\{2,2\mathrm{Re}\lambda ,2\mathrm{Re}\lambda
^{2},0,0,0\}=\{2,-1,-1,0,0,0\}.
\end{equation*}

The character $\chi ^{L}=(6,0,0,0,0,0)$ can be decomposed as
\begin{equation*}
\chi ^{L}=\chi ^{0}+\chi ^{1}+2\chi ^{2}
\end{equation*}%
and therefore the left regular representation is unitarily equivalent to
\begin{equation*}
D^{0}\oplus D^{1}\oplus D^{2}\oplus D^{2}.
\end{equation*}

Examples of positive-type functions are the diagonal elements of any unitary
representation and any linear combination of them with positive coefficient.
Characters are therefore positive-type functions. For instance the Naimark
matrix of $\chi ^{2}$ is
\begin{equation}
N\left( \chi ^{2}\right) =\left[%
\begin{array}{rrrrrr}
2 & -1 & -1 & 0 & 0 & 0 \\
-1 & 2 & -1 & 0 & 0 & 0 \\
-1 & -1 & 2 & 0 & 0 & 0 \\
0 & 0 & 0 & 2 & -1 & -1 \\
0 & 0 & 0 & -1 & 2 & -1 \\
0 & 0 & 0 & -1 & -1 & 2%
\end{array}%
\right]
\end{equation}%
and has eigenvalues: $0,0,3,3,3,3.$

One can write the most general positive-type function on the group. The most
general $N\left( f\right) ,$ which takes care only of Hermiticity
conditions, must be proportional to the matrix%
\begin{equation}
\left[%
\begin{array}{cccccc}
1 & a+\mathrm{i}\ b & a-\mathrm{i}\ b & r & s & t \\
a-\mathrm{i}\ b & 1 & a+\mathrm{i}\ b & t & r & s \\
a+\mathrm{i}\ b & a-\mathrm{i}\ b & 1 & s & t & r \\
r & t & s & 1 & a-\mathrm{i}\ b & a+\mathrm{i}\ b \\
s & r & t & a+\mathrm{i}\ b & 1 & a-\mathrm{i}\ b \\
t & s & r & a-\mathrm{i}\ b & a+\mathrm{i}\ b & 1%
\end{array}%
\right]
\end{equation}%
where $a,b,r,s,t$ are real. The different eigenvalues are%
\begin{equation}
2a+1\pm \left( r+s+t\right) ,1-a\pm \sqrt{3b^{2}-rt-st-rs+r^{2}+s^{2}+t^{2}}.
\end{equation}%
The function
\begin{equation}
f=\left( 1,a+\mathrm{i}b,a-\mathrm{i}b,r,s,t\right)  \label{genpos}
\end{equation}
is of positive-type \textsl{iff} these eigenvalues are nonnegative.

Let us consider the matrix $M$\ constructed by taking as rows the matrix
elements with the same row label in all the irreducible group representation
matrices, normalized to be a unity norm vector. The matrix $M$\ is unitary,
due to the the orthogonality relations $\left( \ref{orthog}\right).$ It reads%
\begin{equation*}
M=\left[%
\begin{array}{rrrrrr}
\frac{1}{\sqrt{6}} & \frac{1}{\sqrt{6}} & \frac{1}{\sqrt{6}} & \frac{1}{%
\sqrt{6}} & \frac{1}{\sqrt{6}} & \frac{1}{\sqrt{6}} \\
\frac{1}{\sqrt{6}} & \frac{1}{\sqrt{6}} & \frac{1}{\sqrt{6}} & -\frac{1}{%
\sqrt{6}} & -\frac{1}{\sqrt{6}} & -\frac{1}{\sqrt{6}} \\
\frac{1}{\sqrt{3}} & \frac{\lambda }{\sqrt{3}} & \frac{\lambda ^{2}}{\sqrt{3}%
} & 0 & 0 & 0 \\
0 & 0 & 0 & \frac{1}{\sqrt{3}} & \frac{\lambda }{\sqrt{3}} & \frac{\lambda
^{2}}{\sqrt{3}} \\
0 & 0 & 0 & \frac{1}{\sqrt{3}} & \frac{\overline{\lambda }}{\sqrt{3}} &
\frac{\overline{\lambda }^{2}}{\sqrt{3}} \\
\frac{1}{\sqrt{3}} & \frac{\overline{\lambda }}{\sqrt{3}} & \frac{\overline{%
\lambda }^{2}}{\sqrt{3}} & 0 & 0 & 0%
\end{array}%
\right] \, .
\end{equation*}%
The matrix $M$ diagonalizes the Naimark matrix of $\chi ^{2}:$

\begin{equation}
M^{\dagger }N\left( \chi ^{2}\right) M=\mathrm{diag}\left[ 0,0,3,3,3,3\right]
\, .
\end{equation}

Using the orthogonality relations, it can be shown that, for any finite or
compact group, the Naimark matrix of characters is diagonalized by the
corresponding $M$ matrix.

Finally, recalling the remark in the end of subsection 4.1, we note that the
above explicit form of the unitary matrix $M$ solves the problem of
determining the unitary elements $f$ $^{^{\prime }}$s in the group algebra
of $S_{3}$ as:%
\begin{equation}
f=M^{\dagger }(u^{0},u^{1},u_{11}^{2},u_{12}^{2},u_{21}^{2},u_{22}^{2})^{%
\mathrm{T}}
\end{equation}%
where $u^{0},u^{1}\in U(1)$ and the matrix $u^{2}$ belongs to $U(2).$

\subsection{The group algebra of a compact Lie group}

The notion of group algebra can be extended to compact Lie groups. The
essential aspect for the definition of a group algebra is the existence of
an (bi--)invariant measure, the Haar measure $dg$. Thus any continuous
function $f:G\rightarrow \mathbb{C}$ on a compact Lie group is integrable
with respect to the Haar measure:%
\begin{equation}
\int\limits_{G}f(g)dg<\infty .
\end{equation}%
and the integral is invariant under left as well as right actions:%
\begin{equation}
\int\limits_{G}f(gh)dg=\int\limits_{G}f(gh)dh=\int\limits_{G}f(g)dg.
\end{equation}%
The measure $dg$ is normalized in such a way that the volume of the group is
one. We will consider the algebra $\mathcal{A}_G$ consisting on all
integrable functions on the group $G$, i.e., $\mathcal{A}_G = L^1(G, dg)$,
together with the convolution product. Thus if $A$ is the element on $%
\mathcal{A}_G$ represented by the function $f_A$, we will have that the
element $A\cdot B$ is represented by the function%
\begin{equation}
\int\limits_{G}f_{A}(h)f_{B}(h^{-1}g)dh =
\int\limits_{G}f_{A}(gh^{-1})f_{B}(h)dh = f_{A\cdot B}(g),
\end{equation}%
along with%
\begin{equation}
\mathrm{Tr}\left[ A^{\dagger }B\right] =\int\limits_{G}f_{A}^{\ast
}(g)f_{B}(g)dg .
\end{equation}%
Other properties of the finite group algebra are extended very easily in
terms of representing functions.

For instance, consider the group $U(1),$ with group manifold the circle $%
0\leq \theta <2\pi .$ The Abelian group $U(1)$ has irreducible
one-dimensional representations labelled by integers:%
\begin{equation}
D^{m}:\theta \mapsto \exp (\mathrm{i}m\theta ),\quad m\in \mathbb{Z} ,
\end{equation}
and their characters are: $\chi ^{m}(\theta )=\exp (\mathrm{i}m\theta ).$

The corresponding $M$ matrix has discrete row and continuous column
labelling indices
\begin{equation}
\left( M_{m\theta }\right) =\frac{1}{\sqrt{2\pi }}\left( \exp (\mathrm{i}%
m\theta )\right) .
\end{equation}%
Of course, it is unitary, that is%
\begin{equation}
\sum_{\theta }\left( M_{m\theta }\right) \left( M_{m^{\prime }\theta }^{\ast
}\right) =\frac{1}{2\pi }\int_{0}^{2\pi }\exp (\mathrm{i}(m-m^{\prime
})\theta )d\theta =\delta _{m,m^{\prime }}
\end{equation}%
and%
\begin{equation}
\sum_{m}\left( M_{m\theta ^{\prime }}^{\ast }\right) \left( M_{m\theta
}\right) =\frac{1}{2\pi }\sum_{m}\exp (\mathrm{i}m(\theta -\theta ^{\prime
}))=\delta (\theta -\theta ^{\prime }).
\end{equation}%
\

The Naimark matrix of $\chi ^{m}$ has elements $\exp \left[ \mathrm{i}%
m\left( \theta ^{\prime }-\theta \right) \right] /2\pi $ and is diagonalized
by $M:$%
\begin{eqnarray}
&&\left( M^{\dagger }N\left( \chi ^{m}\right) M\right) _{m_{1}m_{2}} \\
&=&\frac{1}{4\pi ^{2}}\int_{0}^{2\pi }\int_{0}^{2\pi }\exp \left[ \mathrm{i}%
(m_{1}-m)\theta +\mathrm{i}(m-m_{2})\theta ^{\prime }\right] d\theta d\theta
^{\prime }=\delta _{m_{1},m_{2}}.  \nonumber
\end{eqnarray}

\subsection{States and tomograms in two dimensions}

States in two dimensions are parametrized by points of the 3-dimensional
solid sphere%
\begin{equation}
\rho =\frac{1}{2}\left[
\begin{array}{cc}
1 & 0 \\
0 & 1%
\end{array}%
\right] +x\sigma _{x}+y\sigma _{y}+z\sigma _{z}=\frac{1}{2}\left[
\begin{array}{cc}
1+z & x-\mathrm{i}y \\
x+\mathrm{i}y & 1-z%
\end{array}%
\right]
\end{equation}%
where%
\begin{equation}
\sigma _{x}=\frac{1}{2}\left[
\begin{array}{cc}
0 & 1 \\
1 & 0%
\end{array}%
\right] , \quad \sigma _{y}=\frac{1}{2}\left[
\begin{array}{cc}
0 & -\mathrm{i} \\
\mathrm{i} & 0%
\end{array}%
\right] , \quad \sigma _{z}=\frac{1}{2}\left[
\begin{array}{cc}
1 & 0 \\
0 & -1%
\end{array}%
\right]
\end{equation}%
are the Pauli matrices. The eigenvalues of $\rho $ are%
\begin{equation}
\rho _{\mp }=\frac{1}{2}\left( 1\mp r\right)
\end{equation}%
where by the positivity condition $r^{2}=x^{2}+y^{2}+z^{2}\leq 1$, so $%
(x,y,z)$ is a point of a ball (Bloch sphere) of radius 1 centered at the
origin and the pure states are points on the surface $x^{2}+y^{2}+z^{2}=1.$

The diagonal matrices $d_{g}$'s for $D^{2}$ are
\begin{equation}
\left[
\begin{array}{cc}
1 & 0 \\
0 & 1%
\end{array}%
\right] ,\quad \left[
\begin{array}{cc}
\lambda & 0 \\
0 & \overline{\lambda }%
\end{array}%
\right] , \quad \left[
\begin{array}{cc}
\lambda ^{2} & 0 \\
0 & \overline{\lambda }^{2}%
\end{array}%
\right] , \quad \left[
\begin{array}{cc}
-1 & 0 \\
0 & 1%
\end{array}%
\right] _{j=4,5,6},  \nonumber
\end{equation}%
while the diagonalizing $V_{g}$'s, such that $V_{g}^{\dagger
}D^{2}(g)V_{g}=d_{g},$ are respectively
\begin{equation}
\hskip -1.5cm \left[
\begin{array}{cc}
1 & 0 \\
0 & 1%
\end{array}%
\right] _{j=1,2,3},\frac{1}{\sqrt{2}}\left[
\begin{array}{cc}
-1 & 1 \\
1 & 1%
\end{array}%
\right] ,\frac{1}{\sqrt{2}}\left[%
\begin{array}{cc}
-e^{\mathrm{i}\frac{2\pi }{3}} & e^{\mathrm{i}\frac{2\pi }{3}} \\
1 & 1%
\end{array}%
\right],\frac{1}{\sqrt{2}}\left[%
\begin{array}{cc}
-e^{i\frac{4\pi }{3}} & e^{i\frac{4\pi }{3}} \\
1 & 1%
\end{array}%
\right].  \nonumber
\end{equation}%
The $V_{g}$'s are determined up to phases, one for each column; tomograms $%
(V_{g}^{\dagger }\rho V_{g})_{mm}$ are invariant under the change of these
phases. The first $V_{g}=V_{e}$ is an arbitrary unitary matrix, here chosen
as the identity. At the point $g=e$ the tomogram is an arbitrary stochastic
vector: this is in agrement with the probabilistic interpretation of the
tomogram as probability of getting the eigenvalues of $D(e)$ in a measure.

The matrices $\left\{ V_{g_{j}}^{\dagger }\rho V_{g_{j}}\right\} _{j=1,\dots
,6}$ are%
\begin{eqnarray}
&&V_{g_{j}}^{\dagger }\rho V_{g_{j}}=\frac{1}{2}\left[
\begin{array}{cc}
1+z & x-\mathrm{i}y \\
x+\mathrm{i}y & 1-z%
\end{array}%
\right] ,\quad j=1,2,3 \\
&&V_{g_{j}}^{\dagger }\rho V_{g_{j}}=\frac{1}{2}\left[
\begin{array}{cc}
1-x & -\left( z-\mathrm{i}y\right) \\
-\left( z+\mathrm{i}y\right) & 1+x%
\end{array}%
\right] ,\quad j=4  \nonumber \\
&&V_{g_{j}}^{\dagger }\rho V_{g_{j}}=\frac{1}{2}\left[%
\begin{array}{cc}
1+\frac{1}{2}\left( x+\sqrt{3}y\right) & -z-\frac{1}{2}\mathrm{i}\left( y-%
\sqrt{3}x\right) \\
-z+\frac{1}{2}\mathrm{i}\left( y-\sqrt{3}x\right) & 1-\frac{1}{2}\left( x+%
\sqrt{3}y\right)%
\end{array}%
\right],\quad j=5,  \nonumber \\
&&V_{g_{j}}^{\dagger }\rho V_{g_{j}}=\frac{1}{2}\left[%
\begin{array}{cc}
1+\frac{1}{2}\left( x-\sqrt{3}y\right) & -z-\frac{1}{2}\mathrm{i}\left( y+%
\sqrt{3}x\right) \\
-z+\frac{1}{2}\mathrm{i}\left( y+\sqrt{3}x\right) & 1-\frac{1}{2}\left( x-%
\sqrt{3}y\right)%
\end{array}%
\right],\quad j=6.  \nonumber
\end{eqnarray}

The tomograms \textbf{\ }$(V_{g_{j}}^{\dagger }\rho V_{g_{j}})_{mm}$\textbf{%
\ }for a generic two dimensional state with respect to the representation $%
D^{2}$ are the stochastic vectors%
\begin{equation}
\begin{array}{c}
\frac{1}{2}\left[%
\begin{array}{c}
1+z \\
1-z%
\end{array}%
\right]_{j=1,2,3},\frac{1}{2}\left[%
\begin{array}{c}
1-x \\
1+x%
\end{array}%
\right]_{j=4}, \\
\frac{1}{2}\left[%
\begin{array}{c}
1+\frac{1}{2}\left( x+\sqrt{3}y\right) \\
1-\frac{1}{2}\left( x+\sqrt{3}y\right)%
\end{array}%
\right]_{j=5},\frac{1}{2}\left[%
\begin{array}{c}
1+\frac{1}{2}\left( x-\sqrt{3}y\right) \\
1-\frac{1}{2}\left( x-\sqrt{3}y\right)%
\end{array}%
\right]_{j=6}%
\end{array}%
\end{equation}

\subsection{Positive-type functions}

In view of the Proposition \ref{convex_decomp}, any positive group function
has the form $\varphi _{\rho }=\mathrm{Tr}\left[ \rho D\right] $ , with $%
D=D^{0}\oplus D^{1}\oplus D^{2}$ and $\rho $ decomposes accordingly. The $%
4\times 4-$density state $\rho $ has the form
\begin{equation}
\rho =\left[%
\begin{array}{cccc}
\alpha & 0 & 0 & 0 \\
0 & \beta & 0 & 0 \\
0 & 0 & \frac{1}{2}\gamma \left( 1+z\right) & \frac{1}{2}\gamma \left( x-%
\mathrm{i}y\right) \\
0 & 0 & \frac{1}{2}\gamma \left( x+\mathrm{i}y\right) & \frac{1}{2}\gamma
\left( \frac{1}{2}-z\right)%
\end{array}%
\right],
\end{equation}%
where $\alpha ,\beta ,\gamma \geq 0,$ with$\ \alpha +\beta +\gamma =1,$\
while the positive type function $\varphi _{\rho }=\mathrm{Tr}\left[ \rho D%
\right] $ has values%
\begin{equation}
\varphi _{\rho }=\left[
\begin{array}{c}
1 \\
\alpha +\beta -\frac{1}{2}\gamma \left( 1-\mathrm{i}\sqrt{3}z\right) \\
\alpha +\beta -\frac{1}{2}\gamma \left( 1+\mathrm{i}\sqrt{3}z\right) \\
\alpha -\beta +\gamma x \\
\alpha -\beta -\frac{1}{2}\gamma \left( x+\sqrt{3}y\right) \\
\alpha -\beta -\frac{1}{2}\gamma \left( x-\sqrt{3}y\right)%
\end{array}%
\right] .
\end{equation}%
This vector gives explicitly the form previously obtained in eq. $\left( \ref%
{genpos}\right) .$ For $\alpha ,\beta =0$ this gives positive functions when
only $D^{2}$ is present.

\section{Tomogram associated with $U(n)$ groups}

In this section we introduce the tomograms of states associating the
tomographic probabilities with the group $U(n)$ and other compact Lie groups
$G$. Since $U(n)$ can be factorized as $U(1)\times SU(n)$ up to a quotient by $\mathbb{Z}_n$, we will be mainly
concerned with $SU(n).$

As a general remark, we observe that all the previous results can be
straightforwardly extended to the present case. The diagonalization
procedure leading to the tomographic scheme for finite groups is recovered
by means of the theory of maximal tori for compact groups. In fact, the
diagonalization procedure provides a set of eigen-projectors, containing a
family of rank-one projectors which is tomographic, i.e., it is a resolution
of the identity. In the compact group case, the tomographic family is
obtained by group action on a fiducial set of rank-one projectors, obtained
by the eigenvectors of a complete set of commuting observables.

We begin with a review of some results on compact Lie groups $G$ that will
be needed in what follows \cite{Speiser}.

Any element $g$ of $G$ lies on a one-parameter subgroup $L$ which needs not
be compact, and whose closure is a torus $T$. Every such torus $T$ is
contained in a maximal torus $T_{\max }$, so that any element of the group
belongs to a maximal toroid at least. All maximal tori are conjugated: if $%
T_{\max }$ and $T_{\max }^{\prime }$ \ are maximal tori, there exists an
element $g$ such that $T_{\max }^{\prime }=gT_{\max }g^{-1}$. So, maximal
tori have the same dimension $r,$ the rank of the group $G.$ Besides, $G$
may be obtained by conjugating a fixed maximal torus $T_{\max }$ by all
elements of $G,$ or, more simply, by representative elements $g$ of cosets $%
\left[ g\right] $ of $G/T_{\max }:$%
\begin{equation}
G=\bigcup_{g\in G}gT_{\max }g^{-1}=\bigcup_{\left[ g\right] \in G/T_{\max
}}gT_{\max }g^{-1}  \label{tororbit}
\end{equation}%
An element $t$ of $T_{\max }$ is called regular if it does not belong to any
other maximal torus, otherwise the element is singular. In other words, $t$
is singular if and only if there exists $g\notin $ $T_{\max }$ such that $%
gtg^{-1}=t$, in particular the unity of $G$ is singular.

From a tomographic point of view, it is necessary to describe the previous
results by using the Lie algebra $\mathfrak{g}$ of $G$ and a Cartan
subalgebra $\mathfrak{h}\subset \mathfrak{g}$, which is mapped in a maximal
torus $T_{\mathfrak{h}}$ by the exponential map. To characterize the
singular elements $t$ of $T_{\mathfrak{h}}$ we introduce a basis of $d$
generators of $\mathfrak{g}$: $E_{1},...,E_{d-r},H_{1},...,H_{r}$ . Upon
putting $t=\exp \xi ^{b}H_{b}$ , $g=\exp \eta ^{a}E_{a}$, (hereafter we
adopt Einstein summation convention) we look for solutions of $gtg^{-1}=t$ ,
$g\notin $ $T_{\mathfrak{h}}$ , at the level of Lie algebra, in the form
\begin{equation}
\left[ \eta ^{a}E_{a},\xi ^{b}H_{b}\right] =0.
\end{equation}%
This amounts to
\begin{equation}
C_{a,b}^{a^{\prime }}\xi ^{b}\eta ^{a}=0;a,a^{\prime }=1,...,d-r,
\end{equation}%
where $C_{a,b}^{a^{\prime }}$ are structure constants of the algebra of the
group $G.$ The above square linear system in the unknowns $\left\{ \eta
^{a}\right\} $ yields the commutant, external to the Cartan subalgebra, of
the given element $\xi ^{b}H_{b}$. \ Non-trivial solutions correspond to
singular elements $t=\exp \left( \xi ^{b}H_{b}\right) .$ If the compact Lie
group $G$ is semisimple we can identify its Lie algebra and its dual by
means of the Killing--Cartan form. The dimension of the orbit of the
(co-)adjoint action of the group on its Lie algebra through a singular point
$\xi ^{b}H_{b}$ is smaller than that of the orbit through a regular point,
which is $d-r.$ The same holds for the action of the group on itself by
conjugation. We recall that all the co-adjoint orbits, both regular and
singular, are symplectic manifolds, hence endowed with invariant measures%
\textsl{.} Besides, from a measure theoretical point of view, the set of all
singular orbits has zero Haar measure in the group. As a consequence,
integration of functions on the group may be performed via Fubini's theorem,
integrating over a maximal torus $T_{\mathfrak{h}}$ and the integral over a
regular orbit through $t,$ times a Jacobian taking into account the
dependence on $t. $ Quite generally, this Jacobian can be evaluated for any
compact Lie group \cite{Segal}.

Quantum tomography requires the use of an irreducible unitary group
representation\ $D(g).$ Assume $D$ is the defining representation of $%
G=SU(n).$ Then the Cartan subalgebra generators $\left\{ H_{b}\right\} $
become a complete set of commuting observables of a physical system. From
the previous analysis, we know that the spectrum degeneracy of $\xi
^{b}H_{b} $\ is even for singular points.

The adjoint action on the maximal torus gives rise to the family of unitary
operators $D(g)\exp (\mathrm{i}\xi ^{b}H_{b})D^{\dagger }(g).$ By
decomposing the vector space $\mathfrak{g}=\mathfrak{h}\oplus \mathfrak{e}$
as a direct sum of orthogonal subspaces, and choosing accordingly the basis $%
H_{1},...,H_{r},E_{1},...,E_{d-r},$we observe that the elements $\exp \eta
^{a}E_{a}$ parametrize $G/T_{\mathfrak{h}}$ so that $D(g)\exp (\mathrm{i}\xi
^{b}H_{b})D^{\dagger }(g)$ can be parametrized by $\left( \xi ^{b},\eta
^{a}\right) $, i.e.,%
\begin{equation}
D(g)\exp (\mathrm{i}\xi ^{b}H_{b})D^{\dagger }(g)=D(\tilde{g}),
\end{equation}
where $\tilde{g}=\left( \xi ^{b},\eta ^{a}\right) $ covers almost everywhere
the whole group $G.$ In other words, $D(\tilde{g})$ is diagonalized by $D(g)$
and is iso-spectral with $\exp (\mathrm{i}\xi ^{b}H_{b}).$ Both these
matrices belong to the representation, in contrast with the finite group
case, where the diagonalizing matrix $V_{g_{k}}$ and the diagonal matrix $%
d_{g_{k}}$ do not belong to the representation.

We note that, as $D(g)$ diagonalizes $D(\tilde{g})$ for any $g\in \left[ g%
\right] ,$ one can choose $g=\left( 0,\eta ^{a}\right) $ to avoid
redundancies. The above invariant integration on the group may be performed
according to that parametrization.

By using the projector valued measure (PVM) $\Pi (\xi ^{b})\left( \cdot
\right) $ associated to the Hermitian operator $\xi ^{b}H_{b},$ the spectral
decomposition of $D(\tilde{g})$ may be written as
\begin{equation}
D(\tilde{g})=\int\limits_{\mathbb{R}}\mathrm{e}^{\mathrm{i}\lambda }\exp
\left( \mathrm{i}\eta ^{a}E_{a}\right) \Pi (\xi ^{b})(d\lambda )\exp \left( -%
\mathrm{i}\eta ^{a}E_{a}\right) .
\end{equation}
By means of a density state of a physical system $\rho $ we define a
positive-type group function $\varphi \left( \tilde{g}\right) =\mathrm{Tr}%
\left[ \rho D(\tilde{g})\right] $ in terms of a probability measure $%
\mathcal{M}_{\rho }$:%
\begin{eqnarray}
\varphi \left( \tilde{g}\right) &=&\int\limits_{\mathbb{R}}\mathrm{e}^{%
\mathrm{i}\lambda }\mathrm{Tr}\left[ \rho \exp \left( \mathrm{i}\eta
^{a}E_{a}\right) \Pi (\xi ^{b})(d\lambda )\exp \left( -\mathrm{i}\eta
^{a}E_{a}\right) \right]  \nonumber \\
&=&\int\limits_{\mathbb{R}}\mathrm{e}^{\mathrm{i}\lambda }\mathcal{M}_{\rho
}(\xi ^{b},\eta ^{a})(d\lambda ).
\end{eqnarray}%
The probability measure $\mathcal{M}_{\rho }$, which is labelled by $\tilde{g%
}=\left( \xi ^{b},\eta ^{a}\right) ,$ is related to the tomogram associated
to the density state $\rho ,$ in the tomographic scheme based on the group $%
G.$

More precisely,
\begin{equation}
\int\limits_{B}\mathcal{M}_{\rho }(\xi ^{b},\eta ^{a})(d\lambda )
\end{equation}%
is the probability that a measure of the observable $\xi ^{b}H_{b}$ in the
rotated state $\exp \left( -\mathrm{i}\eta ^{a}E_{a}\right) \rho \exp \left(
\mathrm{i}\eta ^{a}E_{a}\right) $ belongs to the Borel set $B$ of the real
line. As a consequence:
\begin{equation}
\int\limits_{B}\mathcal{M}_{\rho }(\xi ^{b},\eta ^{a})(d\lambda
)=\int\limits_{kB}\mathcal{M}_{\rho }(k\xi ^{b},\eta ^{a})(d\lambda ^{\prime
})=\int\limits_{B}\left\vert k\right\vert \mathcal{M}_{\rho }(k\xi ^{b},\eta
^{a})(d\lambda )
\end{equation}%
and we get the homogeneity property
\begin{equation}
\mathcal{M}_{\rho }(k\xi ^{b},\eta ^{a};d\lambda )=\frac{1}{\left\vert
k\right\vert }\mathcal{M}_{\rho }(\xi ^{b},\eta ^{a};d\lambda ).  \label{hom}
\end{equation}

In view of the compactness of $G,$ all the unitary irreducible
representations (UIR's) are finite dimensional and the PVM of $\xi ^{b}H_{b}$
is concentrated on a set of $n=\mathrm{\dim }D$ points $\left\{ \mu
_{m}\right\} _{m},\mu _{m}=\xi ^{b}m_{b}$ where $m_{b}$ is an eigenvalue of $%
H_{b}, b=1,...,r$, while $m=1,...,n:$
\begin{equation}
\Pi (\xi ^{b})(d\lambda )=\sum\limits_{\left\{ m_{b}\right\} }P_{\left\{
m_{b}\right\} }\delta (\lambda -\xi ^{b}m_{b})d\lambda ,
\end{equation}%
and where a Gelfand-Zetlin basis has been chosen in such a way that the
rank-one projector $P_{\left\{ m_{b}\right\} }$ projects on the eigenspace
of the eigenvalue $\xi ^{b}m_{b},$ which is the same eigenspace of the
eigenvalues $m_{b}$, for any $b.$ Then we can define the tomogram of the
state $\rho ,{W}_{\rho }(\eta ^{a}\,;m),$ with respect to the representation
$D$ of the group $G$:
\begin{eqnarray}
\mathrm{Tr}(\rho D(\tilde{g})) &=&\sum\limits_{\left\{ m_{b}\right\} }\exp (%
\mathrm{i}\xi ^{b}m_{b})\mathrm{Tr}\left[ \exp \left( -\mathrm{i}\eta
^{a}E_{a}\right) \rho \exp \left( \mathrm{i}\eta ^{a}E_{a}\right) P_{\left\{
m_{b}\right\} }\right] \\
&=&\sum\limits_{\left\{ m_{b}\right\} }\exp (\mathrm{i}\xi ^{b}m_{b})\mathrm{%
Tr}\left[ \rho \exp \left( \mathrm{i}\eta ^{a}E_{a}\right) P_{\left\{
m_{b}\right\} }\exp \left( -\mathrm{i}\eta ^{a}E_{a}\right) \right]  \nonumber
\\
&=&\sum\limits_{\left\{ m_{b}\right\} }\exp (\mathrm{i}\xi ^{b}m_{b}){W}%
_{\rho }(\eta ^{a}\,;\left\{ m_{b}\right\} ).
\end{eqnarray}

In other words, the tomogram $\left\{ {W}_{\rho }(\eta ^{a}\,;\left\{
m_{b}\right\} )\right\} $ is a stochastic vector:%
\begin{equation}
\sum\limits_{\left\{ m_{b}\right\} }{W}_{\rho }(\eta ^{a}\,;\left\{
m_{b}\right\} )=1.
\end{equation}%
The component ${W}_{\rho }(\eta ^{a}\,;\left\{ m_{b}\right\} )$ is the joint
probability that a measure of any $H_{b}$ in the rotated state $\tilde{\rho}%
=\exp \left( -i\eta ^{a}E_{a}\right) \rho \exp \left( i\eta ^{a}E_{a}\right)
$ is $m_{b}$ respectively:%
\begin{eqnarray}
\mathrm{Tr}[\tilde{\rho}H_{b^{\prime }}] &=& \mathrm{Tr}\left[ \exp \left( -%
\mathrm{i}\eta ^{a}E_{a}\right) \rho \exp \left( \mathrm{i}\eta
^{a}E_{a}\right) \sum\limits_{\left\{ m_{b}\right\} }m_{b^{\prime
}}P_{\left\{ m_{b}\right\} }\right]  \nonumber \\
& = & \sum\limits_{\left\{ m_{b}\right\} }m_{b^{\prime }}{W}_{\rho }(\eta
^{a}\,;\left\{ m_{b}\right\} )
\end{eqnarray}

We observe explicitly that the tomogram can be viewed equivalently as a
measure of the rotated observable $\exp \left( \mathrm{i}\eta
^{a}E_{a}\right) H_{b^{\prime }}\exp \left( -\mathrm{i}\eta ^{a}E_{a}\right)
$ in the state $\rho .$ In other words, out of the fiducial set of rank one
projectors $P_{\left\{ m_{b}\right\} }$'s, one gets a tomographic set of
rotated rank-one projectors. Of course, the density state $\rho $ can be
reconstructed from its tomogram ${W}_{\rho }.$ To this aim, we observe that
from the orthogonality relations we get
\begin{equation}
d^{\left( D\right) }\int\limits_{G}\varphi (\tilde{g})^{\ast }D(\tilde{g})d%
\tilde{g}=d^{\left( D\right) }\int\limits_{G}\mathrm{Tr}\left[ \rho D(\tilde{%
g})\right] ^{\ast }D(\tilde{g})d\tilde{g}=\rho ,  \label{compRec}
\end{equation}%
where $d^{\left( D\right) }$, the formal dimension, is the dimension of $D$
divided by the Haar volume of the group. That is, taking into account the
reality of the tomograms,%
\begin{equation}
d^{\left( D\right) }\int\limits_{G}\sum\limits_{\left\{ m_{b}\right\} }\exp
(-\mathrm{i}\xi ^{b}m_{b}){W}_{\rho }(\eta ^{a}\,;\left\{ m_{b}\right\} )D(%
\tilde{g})d\tilde{g}=\rho .
\end{equation}

We observe that the above equation may be further detailed in particular
cases.

For instance, in the $SU\left( 2\right) $ case with $D=D^{j}$ of $2j+1$
dimensions. We preliminarly note that, in general, as $\tilde{g}=gtg^{-1}$
with $t\in T$ and $g\in G,$ for any summable group function $f$:%
\begin{equation}
\int\limits_{G}f\left( \tilde{g}\right) \mu _{G}\left( d\tilde{g}\right)
=\int\limits_{T}\int\limits_{G}f\left( gtg^{-1}\right) \mu _{G}\left(
dg\right) \mu _{T}\left( dt\right)
\end{equation}
where $\mu _{G}$ and$\ \mu _{T}$ are normalized invariant measure on $G$ and
$T$ respectively.

Then, in the canonical basis of the eigenvectors $\left\{ \left\vert
m\right\rangle \right\} ,m=-j,\dots ,j,$ of $J_{z}$ we have:
\begin{equation*}
\hskip -2cm \rho _{m_{1}m_{2}} = d^{\left(
D\right)}\int\limits_{G}\sum\limits_{m=-j}^{j}\exp (-\mathrm{i}\xi m){W}%
_{\rho }(\eta ^{a};m)D_{m_{1}m_{2}}(\tilde{g})d\tilde{g} =
\end{equation*}
\begin{equation*}
\hskip -2cm = \frac{d^{\left( D\right) }}{2\pi }\int\limits_{0}^{2\pi }d\xi
\int\limits_{G}\sum\limits_{m,m^{\prime }=-j}^{j}\exp [\mathrm{i}\xi
(m^{\prime }-m)]{W}_{\rho }(g\,;m)\left( D\left( g\right) \left\vert
m^{\prime }\right\rangle \left\langle m^{\prime }\right\vert D^{\dagger
}\left( g\right) \right) _{m_{1}m_{2}}dg =
\end{equation*}
\begin{equation}
= d^{\left( D\right) }\sum\limits_{m=-j}^{j}\int\limits_{G}{W}_{\rho
}(g\,;m)\left( D\left( g\right) \left\vert m\right\rangle \left\langle
m\right\vert D^{\dagger }\left( g\right) \right) _{m_{1}m_{2}}dg.
\end{equation}

The expression $D\left( g\right) \left\vert m\right\rangle \left\langle
m\right\vert D^{\dagger }\left( g\right) $ is just the action of the group
on $\mathcal{H}\otimes \mathcal{H}^{\ast },$ where $\mathcal{H}$ is the
carrier space of $D$ and \ $\mathcal{H}^{\ast }$ its dual, the carrier space
of the transpose representation $D^{T}(g^{-1}):D^{T}(g^{-1})\left( m,\cdot
\right) =\left( m,D(g^{-1})\cdot \right) =\left\langle m\right\vert
D^{\dagger }\left( g\right) .$

Now, for $SU(2),$ the representations $D\left( g\right) $ and its complex
conjugate $D^{\ast }(g)=D^{T}(g^{-1})$ are equivalent for any $j,$ so that
we can use the contravariant basis (\cite{Lyubarskii}, sec. 41): $%
\left\langle m\right\vert \mapsto \left( -1\right) ^{j-m}\left\vert
-m\right\rangle ,$ in such a way that the group action on $\mathcal{H}%
\otimes \mathcal{H}^{\ast }$ is equivariant with the group action on $%
\mathcal{H}\otimes \mathcal{H}$. This allows to use the group action $%
D\otimes D$ on $\mathcal{H}\otimes \mathcal{H}$ and the addition theorem to
decompose the product representation:
\begin{equation}
D^{j}\otimes D^{j}=\bigoplus\limits_{J=0}^{2j}D^{J}.
\end{equation}

Finally, the reconstruction formula for the matrix element $\rho
_{m_{1}m_{2}}$\ reads%
\begin{eqnarray}
& & \rho _{m_{1}m_{2}} = \sum\limits_{m=-j}^{j}d^{\left( D\right)
}\int\limits_{G}{W}_{\rho }(g\,;m)\left( D\left( g\right) \left\vert
m\right\rangle \left\langle m\right\vert D^{\dagger }\left( g\right) \right)
_{m_{1}m_{2}}dg =  \nonumber \\
& = &\sum\limits_{m=-j}^{j}\sum\limits_{J=0}^{2j}\sum\limits_{M,M^{\prime
}=-J}^{J}\left( -1\right) ^{2j-m-m_{2}} d^{\left( D\right) }\int\limits_{G}{W%
}_{\rho }(g\,;m)\left\langle m_{1}\right\vert \left\langle
-m_{2}|JM\right\rangle  \nonumber \\
& &\times \left\langle JM\right\vert D^{J}\left( g\right) \left\vert
JM^{\prime }\right\rangle \left\langle JM^{\prime }|m\right\rangle
\left\vert -m\right\rangle dg =  \nonumber \\
&=&\sum\limits_{m=-j}^{j}\sum\limits_{J=0}^{2j}\sum\limits_{M=-J}^{J}\left(
-1\right) ^{2j-M-m-m_{2}}(2J+1)\left[%
\begin{array}{ccc}
j & j & J \\
m_{1} & -m_{2} & -M%
\end{array}%
\right] \times  \nonumber \\
& & \times \left[%
\begin{array}{ccc}
j & j & J \\
m & -m & 0%
\end{array}%
\right] d^{\left( D\right) }\int\limits_{G}{W}_{\rho
}(g\,;m)D_{M0}^{J}\left( g\right) dg.  \label{MatElRec}
\end{eqnarray}%
where the Wigner $3j-$symbols are introduced.

The above equation may be related to the reconstruction formulae contained
in \cite{BeppeOlgaPhysScripta}.

In fact, by observing that
\begin{eqnarray}
& &{W}_{\rho }(g\,;m) = {W}_{\rho }^{\ast }(g\,;m)=\sum_{m_{1}^{\prime
},m_{2}^{\prime }=-j}^{j}\rho _{m_{1}^{\prime }m_{2}^{\prime }}^{\ast
}\left( D\left( g\right) \left\vert m\right\rangle \left\langle m\right\vert
D^{\dagger }\left( g\right) \right) _{m_{2}^{\prime }m_{1}^{\prime }}^{\ast
} =  \nonumber \\
&=&\sum_{m_{1}^{\prime },m_{2}^{\prime }=-j}^{j}\rho _{m_{1}^{\prime
}m_{2}^{\prime }}^{\ast }\sum\limits_{J^{\prime
}=0}^{2j}\sum\limits_{M^{\prime }=-J^{\prime }}^{J^{\prime }}\left[ \left(
-1\right) ^{2j-M^{\prime }-m-m_{1}^{\prime }}D_{M^{\prime }0}^{J^{\prime
}}\left( g\right) \right] ^{\ast }\times  \nonumber \\
&&\times (2J+1)\left[%
\begin{array}{ccc}
j & j & J^{\prime } \\
m_{2}^{\prime } & -m_{1}^{\prime } & -M^{\prime }%
\end{array}%
\right]\left[%
\begin{array}{ccc}
j & j & J^{\prime } \\
m & -m & 0%
\end{array}%
\right],
\end{eqnarray}%
the integration over the group yields $\delta _{J,J^{\prime }}\delta
_{M,M^{\prime }}.$ By means of the well known identities%
\begin{equation}
(2J+1)\sum\limits_{m=-j}^{j}\left[%
\begin{array}{ccc}
j & j & J \\
m & -m & 0%
\end{array}%
\right]\left[%
\begin{array}{ccc}
j & j & J \\
m & -m & 0%
\end{array}%
\right]= 1,
\end{equation}
\begin{equation*}
\hskip -1cm \sum\limits_{J=0}^{2j}\sum\limits_{M=-J}^{J}(2J+1)\left[%
\begin{array}{ccc}
j & j & J \\
m_{2}^{\prime } & -m_{1}^{\prime } & -M%
\end{array}%
\right]\left[%
\begin{array}{ccc}
j & j & J \\
m_{1} & -m_{2} & -M%
\end{array}%
\right]= \delta _{m_{1},m_{2}^{\prime }}\delta _{m_{2},m_{1}^{\prime }},
\nonumber
\end{equation*}%
the l.h.s. of eq. $\left( \ref{MatElRec}\right) $ eventually gives $\rho
_{m_{2}m_{1}}^{\ast }=\rho _{m_{1}m_{2}}.$

In the general $U\left( n\right) $ case, when the used representation and
its conjugate are equivalent, one can try to follow the previous route to
perform the reconstruction.

However, we note that $SU(2)$ can be embedded irreducibly in the defining
representation of $U(n),$ for any $n.$ So, the above reconstruction formula
is general and can be used to reconstruct density states out of the
restriction of the $U(n)$ tomograms to the subgroup $SU(2).$

Back to the general analysis, we remark that as $\varphi (\tilde{g})=\mathrm{%
Tr}\left[ \rho D(\tilde{g})\right] $ is a function on the group $G$ of
positive type, the theorem of Naimark \cite{Naimark} states that there exist
a unitary representation ${U}$ on a Hilbert space determined by a GNS
construction and a cyclic vector $\psi _{0}$ such that
\begin{equation}
\varphi (g)=\left( \psi _{0},{U}(g)\psi _{0}\right) .
\end{equation}

As a result, following a procedure similar to that discussed in sec.6, if $%
\rho $ is a pure state $\left\vert \psi \right\rangle \left\langle \psi
\right\vert $, then ${U}$ and $\psi _{0}$ are unitarily equivalent to $D$
and $\psi $ respectively. When $\rho $ is a mixed state of rank $r$, then ${U%
}$ is reducible, and can be put in block form of $r$ blocks $\mathcal{V}$
unitarily equivalent to $D:\mathcal{V}(g)=VD(g)V^{\dagger }.$ Then $\rho $
can be reconstructed by
\begin{equation}
d^{\left( U\right) }\int\limits_{G}\varphi ^{\ast }(g)\mathcal{V}(g)dg=V\rho
V^{\dagger }.
\end{equation}%
This extends Proposition \ref{reducible} of Sec. 6 to the compact group case.

Also Proposition \ref{convex_decomp} of Sec. 6 can be extended to the
present case. However, we remark that when an arbitrary irreducible
representation has been chosen instead of the defining one, the Cartan
subalgebra operators are not a complete set any further, and a
Gelfand-Zetlin \cite{GelfandZetlin,Baird-Biede} basis has to be determined
by considering a set of Casimir operators of subgroups: for instance, in the
$SU(3)$ case, the isotopic spin operator. In the next section, we present a
discussion of Gelfand-Zetlin basis construction making contact with
tomographic representations. In fact, tomograms depend not only on the group
parameters, playing the role of ``positions" in configuration space, but
also on Gelfand-Zetlin basis labels, that play the role of ``conjugate
momenta".

\section{Gelfand-Zetlin bases}

Let us comment first on how the tomograms constructed using a unitary
representation of a group $G$ are connected not only with the group itself
but also with the choice of the chain of the subgroups of the group which is
used to determine the basis in the Hilbert space on which is acting the
irreducible representation of the group. In fact, the tomogram $W^{\alpha
}\left( g,m\right) $ is a function of the group element $g,$ of the Casimir
label of the representation $\alpha $ and of the collective label $m$ which
determines \ the basis vector in the corresponding Hilbert space. We remind
how this label $m$ is determined. For example for the $SU\left( 2\right) -$%
group the natural choice of the parameter $m $ is the spin projection on $z-$%
axis for a given value $\alpha =J$ of the Casimir operator $\boldsymbol{J}%
^{2}.$

In a purely group-theoretical formalism that does not use any ``physical"
interpretation of the index $m$ (and index $\alpha $ as angular momentum $J$%
) the basis is determined by the Lie algebra generator $J_{z}$ of the
subgroup $U\left( 1\right) $ of the group $SU\left( 2\right) :$ \ one has
the chain $SU\left( 2\right) \supset U\left( 1\right) .$ In the case of $%
SU\left( 3\right) $ the Gelfand-Zetlin basis is determined by the chain $%
SU\left( 3\right) \supset SU\left( 2\right) \supset U\left( 1\right) $\ of
subgroups embedded into $SU\left( 3\right) .$ In fact one determines the
basis using first the Casimir operators of $SU\left( 3\right) ,$ then using
the Casimir operator of $SU\left( 2\right) $ (corresponding to the value of
the isotopic spin $\boldsymbol{T}^{2}$) and the generators of the Cartan
subalgebra providing the weights $m_{1},m_{2}.$ Due to multiplicity of the
weights, to label Cartan generators eigenvectors one needs the Casimir
operator $\boldsymbol{T}^{2}$ of the subgroup $SU\left( 2\right) $ embedded
into the initial group $SU\left( 3\right) .$ For any higher group $SU\left(
n\right) ,$ the Gelfand-Zetlin basis is constructed by using the chain $%
SU\left( n\right) \supset SU\left( n-1\right) \supset \dots \supset U\left(
1\right) $ of embedded subgroups.

But there exist other possibilities to use different chains of subgroups
embedded into the initial group $G.$ For example one can construct the basis
for the irreducible representations of the group $SU\left( 6\right) $ by
using the subgroup $SU\left( 3\right) \otimes SU\left( 2\right) $ embedded
into $SU\left( 6\right) .$ The basis obtained in this way provides the
possibility to get "quantum numbers" corresponding to standard spins (i.e.,
associated with the group $SU\left( 2\right) $) and the charges associated
with the group $SU\left( 3\right) .$ In fact, the ambiguity in the choice of
the subgroup chains determining the basis index $m$ corresponds to the
ambiguity in the choice of the complete set of commuting observables,
operators acting on the Hilbert space of the irreducible representation of
the group $G.$ Of course, the basis vectors $\left\{ \left\vert \alpha
,m\right\rangle \right\} $ and $\left\{ \left\vert \alpha ,m^{\prime
}\right\rangle \right\} $ determined by Casimirs $\alpha $ and quantum
numbers $m$ and $m^{\prime },$ associated with two different chains of
subgroups embedded into the group $G,$ or with two different complete sets
of commuting observables, are related by a unitary transformation $U:$ $%
\left\vert \alpha ,m\right\rangle =U\left\vert \alpha ,m^{\prime
}\right\rangle .$ In terms of the corresponding rank-one projectors this
reads: $P_{\left\{ \alpha ,m\right\} }=UP_{\left\{ \alpha ,m^{\prime
}\right\} }U^{\dagger }.$

From the tomographic point of view, the role played by $U$ in relating
tomograms associated with different chains of embedded subgroups is the
following. Recalling the definition of tomogram of the density state $\rho ,$
we obtain:
\begin{equation}
W_{\rho }^{\alpha }\left( g,m\right) =W_{U^{\dagger }\rho U}^{\alpha }\left(
g,m^{\prime }\right) .
\end{equation}%
In other words, the tomogram of the density state $\rho $ in the basis $%
\left\{ \left\vert \alpha ,m\right\rangle \right\} $ with respect to the
representation $D^{\alpha }$ is just the tomogram of the transformed density
state $U^{\dagger }\rho U$ in the transformed basis $\left\{ \left\vert
\alpha ,m^{\prime }\right\rangle =U^{\dagger }\left\vert \alpha
,m\right\rangle \right\} $ with respect to the transformed representatin $%
U^{\dagger }D^{\alpha }U.$

\section{ The paradigmatic case of $SU(3)$}

We illustrate the previuos analysis by considering the paradigmatic example
of the group $SU(3).$

The basis vector of irreducible representations of $SU(3)$ are labelled by
the eigenvalues $C_{1}$ and $C_{2}$ of the Casimir operators $\hat{C}_{1}$
and $\hat{C}_{2}.$These in the case of the $SU(2)$ group reduce to the spin
Casimir operator $\boldsymbol{J}^{2}$ with eigenvalues $j(j+1),$ $%
j=0,1/2,... $. After fixing the representation $D^{\left( C_{1},C_{2}\right)
}$ by a pair $\left( C_{1},C_{2}\right) ,$ there is a Gelfand-Zetlin basis $%
\left\{ \left\vert m_{1},m_{2};m_{3}\right\rangle \right\} $\ of the Hilbert
space acted upon by $D^{\left( C_{1},C_{2}\right) }$, labelled by three
quantum numbers $m_{1},m_{2}$ and $m_{3}.$

The quantum numbers $m_{1},m_{2}$ are the spectra of the Cartan subalgebra $%
\left\{ H_{a}\right\} $ operators, i.e.
\begin{equation}
H_{a}\left\vert m_{1},m_{2};m_{3}\right\rangle =m_{a}\left\vert
m_{1},m_{2};m_{3}\right\rangle ,\quad a=1,2  \label{A}
\end{equation}%
and $m_{3}$ is eigenvalue of the Casimir operator $\boldsymbol{T}^{2}$
associated with the $SU(2)$ (isotopic spin) subgroup of $SU(3)$%
\begin{equation}
\boldsymbol{T}^{2}\left\vert m_{1},m_{2},m_{3}\right\rangle
=m_{3}(m_{3}+1)\left\vert m_{1},m_{2},m_{3}\right\rangle  \label{B}
\end{equation}

Let us rotate the basis $\left\vert m_{1},m_{2};m_{3}\right\rangle $ by
applying the representation matrix $D(g)$ of $SU(3)$.We get a new basis%
\begin{equation}
\left\vert m_{1},m_{2};m_{3};g\right\rangle :=D(g)\left\vert
m_{1},m_{2};m_{3}\right\rangle .  \label{C}
\end{equation}%
Then we consider for a group element $\tilde{g}$ the mean value of $D(\tilde{%
g})$in the density state $\rho $ belonging to the Hilbert space of the
irreducible representation; in other words we get the Naimark positive
function%
\begin{equation}
\varphi ^{\left( C_{1},C_{2}\right) }(\tilde{g})=\mathrm{Tr}[\rho D^{\left(
C_{1},C_{2}\right) }(\tilde{g})].
\end{equation}%
Here, dropping the label $\left( C_{1},C_{2}\right) ,$
\begin{equation}
D(\tilde{g})=D(g)D(t)D^{\dagger }(g),\quad D(t)=\exp \left[ \mathrm{i}\left(
\xi ^{1}H_{1}+\xi ^{2}H_{2}\right) \right] .
\end{equation}

Consider the standard spectral decomposition of the unitary matrix $D(\tilde{%
g}):$
\begin{eqnarray}
D(\tilde{g}) &=&\sum\limits_{m_{1},m_{2},m_{3}}\mathrm{e}^{\mathrm{i}\left(
\xi ^{1}m_{1}+\xi ^{2}m_{2}\right) }\left\vert
m_{1},m_{2},m_{3};g\right\rangle \left\langle m_{1},m_{2},m_{3};g\right\vert
\nonumber \\
&=&\sum\limits_{m_{1},m_{2},m_{3}}\mathrm{e}^{\mathrm{i}\left( \xi
^{1}m_{1}+\xi ^{2}m_{2}\right) }P_{m_{1},m_{2},m_{3}}\left( g\right) ,
\end{eqnarray}%
where $P_{m_{1},m_{2},m_{3}}\left( g\right) $ is the rank-one
projector.corresponding to $\left\vert m_{1},m_{2},m_{3};g\right\rangle $.
In other words $\left\{ P_{m_{1},m_{2},m_{3}}\left( g\right) \right\} $ is
the PVM of the observable $\xi ^{1}H_{1}+\xi ^{2}H_{2}$, which is a
concentrated measure on the points $\left\{ \xi ^{1}m_{1}+\xi
^{2}m_{2}\right\} \subset \mathbb{R}$ . We get an expression for the
positive function in the form
\begin{eqnarray}
\varphi _{\rho }(\tilde{g}) &=&\sum\limits_{m_{1},m_{2},m_{3}}\mathrm{e}^{%
\mathrm{i}\left( \xi ^{1}m_{1}+\xi ^{2}m_{2}\right) }\mathrm{Tr}[\rho
P_{m_{1},m_{2},m_{3}}\left( g\right) ] \\
&=:&\sum\limits_{m_{1},m_{2},m_{3}}\mathrm{e}^{\mathrm{i}\left( \xi
^{1}m_{1}+\xi ^{2}m_{2}\right) }W_{\rho }(m_{1},m_{2},m_{3};g)  \nonumber
\end{eqnarray}%
Here we have defined the tomogram of $\rho $ in the irreducible
representation $D^{\left( C_{1},C_{2}\right) }$ of $SU\left( 3\right) $ as
the function $W_{\rho }(m_{1},m_{2},m_{3};g)=\mathrm{Tr}[\rho
P_{m_{1},m_{2},m_{3}}\left( g\right) ].$

\section{Inverse tomographic problem}

Consider an operator $A,$ acting on the same Hilbert space of the
irreducible representation $D^{\alpha }$ of the finite group $G_{K}$. Using
the tomographic symbols $\left\{ \left( V_{g_{j}}^{\alpha \dagger
}AV_{g_{j}}^{\alpha }\right) _{mm}\right\} _{j=1}^{K}$\ of the operator $A,$
the formula holds:%
\begin{eqnarray}
&&\frac{n_{\alpha }}{K}\sum\limits_{j=1}^{K}\sum\limits_{m=1}^{n_{\alpha }}%
\mathrm{e}^{-\mathrm{i}\theta _{m}^{\alpha }\left( g_{j}\right) }\left(
V_{g_{j}}^{\alpha \dagger }AV_{g_{j}}^{\alpha }\right) _{mm}^{\ast
}D_{rs}^{\alpha }\left( g_{j}\right) \\
&=&\frac{n_{\alpha }}{K}\sum\limits_{j=1}^{K}\mathrm{Tr}\left[ AD^{\alpha
}\left( g_{j}\right) \right] ^{\ast }D_{rs}^{\alpha }\left( g_{j}\right)
\nonumber  \label{recform} \\
&=&\sum\limits_{p,q}A_{pq}^{\ast }\frac{n_{\alpha }}{K}\sum%
\limits_{j=1}^{K}D^{\alpha }\left( g_{j}\right) _{qp}^{\ast }D_{rs}^{\alpha
}\left( g_{j}\right) =A_{sr}^{\ast }=A_{rs}^{\dagger }
\end{eqnarray}%
When the operator $A$ is an observable,i.e., $A^{\dagger }=A,$ the equation
above is a reconstruction formula for $A.$

Let us consider the family of $n_{\alpha }-$dimensional vectors
\begin{equation}
\left\{ v_{m}^{\alpha }\left( g_{j}\right) \right\} =\left\{ \left(
V_{g_{j}}^{\alpha \dagger }AV_{g_{j}}^{\alpha }\right) _{mm}\right\}
,j=1,\dots ,K.
\end{equation}%
In view of the above formula, as $\left( V_{g_{j}}^{\alpha \dagger
}AV_{g_{j}}^{\alpha }\right) _{mm}^{\ast }=\left( V_{g_{j}}^{\alpha \dagger
}A^{\dagger }V_{g_{j}}^{\alpha }\right) _{mm},$ they satisfy the
self-consistency relation written in terms of a reproducing kernel $%
R_{pm}^{\alpha }\left( g_{j},g_{h}\right) :$%
\begin{eqnarray}
&&\frac{n_{\alpha }}{K}\sum\limits_{j=1}^{K}\sum\limits_{m=1}^{n_{\alpha
}}R_{pm}^{\alpha }\left( g_{j},g_{h}\right) v_{m}^{\alpha \ast
}(g_{j})=v_{p}^{\alpha \ast }(g_{h}),  \label{selfconsist} \\
&&R_{pm}^{\alpha }\left( g_{j},g_{h}\right) :=\mathrm{e}^{-\mathrm{i}\theta
_{m}^{\alpha }\left( g_{j}\right) }\sum\limits_{r,s=1}^{n_{\alpha }}\left(
V_{g_{h}}^{\alpha \dagger }\right) _{pr}D_{rs}^{\alpha }\left( g_{j}\right)
\left( V_{g_{h}}^{\alpha }\right) _{sp}.  \nonumber
\end{eqnarray}

The vectors $\mathbf{v}^{\alpha }\left( g_{j}\right)$ can be chosen as
stochastic vectors only if $A$\textsl{\ }is a positive semidefinite
observable: i.e., a density state $\rho $, after normalization.

In fact, after diagonalization $A=U\mathrm{diag}\left[ \lambda _{1},\lambda
_{2},\dots ,\lambda _{n_{\alpha }}\right] U^{\dagger },$ we may choose the
arbitrary diagonalizing matrix $V_{e}$ associated to the neutral element $e$
of the group to be $U,V_{e}=U.\ $If $A$ is diagonal we choose the identity
matrix as$\ V_{e}.$ In this way, we get as corresponding column vector just $%
\left( \lambda _{1},\lambda _{2},\dots ,\lambda _{n_{\alpha }}\right) ^{%
\mathrm{T}},$ which is a (normalizable) stochastic vector only when\textsl{\
}all the eigenvalues are nonnegative.

However, the above condition is by no means sufficient: a family of
stochastic vectors can be associated to any observable $A.$

For instance, in the triangle group case, consider the tomographic symbols
of the observable $A:$
\begin{equation}
\hskip -1.5cm A=U\left[%
\begin{array}{cc}
\lambda _{1} & 0 \\
0 & -\lambda _{2}%
\end{array}%
\right]U^{\dagger }=\left[%
\begin{array}{cc}
\lambda _{1}\cos ^{2}\frac{\theta }{2}-\lambda _{2}\sin ^{2}\frac{\theta }{2}
& -\frac{\lambda _{1}+\lambda _{2}}{2}\sin \theta \mathrm{e}^{\mathrm{i}\phi
} \\
-\frac{\lambda _{1}+\lambda _{2}}{2}\sin \theta \mathrm{e}^{-\mathrm{i}\phi }
& \lambda _{1}\sin ^{2}\frac{\theta }{2}-\lambda _{2}\cos ^{2}\frac{\theta }{%
2}%
\end{array}%
\right],
\end{equation}%
where%
\begin{equation}
\lambda _{1},\lambda _{2}>0,\quad U=\left[%
\begin{array}{cc}
\cos \frac{\theta }{2}\mathrm{e}^{\mathrm{i}\frac{\phi +\psi }{2}} & \sin
\frac{\theta }{2}\mathrm{e}^{\mathrm{i}\frac{\phi -\psi }{2}} \\
-\sin \frac{\theta }{2}\mathrm{e}^{-\mathrm{i}\frac{\phi -\psi }{2}} & \cos
\frac{\theta }{2}\mathrm{e}^{-\mathrm{i}\frac{\phi +\psi }{2}}%
\end{array}%
\right].
\end{equation}

The tomographic symbols $\left\{ \left( V_{g_{j}}^{\alpha \dagger
}AV_{g_{j}}^{\alpha }\right) _{mm}\right\} _{j=1}^{K}$ of $A$ are in
consequence%
\begin{eqnarray}
&& \left\{ \left[%
\begin{array}{c}
\lambda _{1}\cos ^{2}\frac{\theta }{2}-\lambda _{2}\sin ^{2}\frac{\theta }{2}
\\
\lambda _{1}\sin ^{2}\frac{\theta }{2}-\lambda _{2}\cos ^{2}\frac{\theta }{2}%
\end{array}%
\right]\right\} _{j=1,2,3},  \nonumber \\
&& \left\{ \frac{1}{2}\left[%
\begin{array}{c}
\lambda _{1}-\lambda _{2}+\left( \lambda _{1}+\lambda _{2}\right) \cos
\left( \phi +\alpha \right) \sin \theta \\
\lambda _{1}-\lambda _{2}-\left( \lambda _{1}+\lambda _{2}\right) \cos
\left( \phi +\alpha \right) \sin \theta%
\end{array}%
\right]\right\} _{\alpha =0,\frac{2\pi }{3},\frac{4\pi }{3}}
\end{eqnarray}%
Picking $\theta =\pi /2,\quad \lambda _{1}-\lambda _{2}=1,$ we have the
vectors
\begin{equation}
\frac{1}{2}\left[%
\begin{array}{c}
1 \\
1%
\end{array}%
\right] , \quad \quad \frac{1}{2}\left[%
\begin{array}{c}
1+\left( 1+2\lambda _{2}\right) \cos \left( \phi +\alpha \right) \\
1-\left( 1+2\lambda _{2}\right) \cos \left( \phi +\alpha \right)%
\end{array}%
\right]_{\alpha =0,\frac{2\pi }{3},\frac{4\pi }{3}}
\end{equation}%
so that, putting $M\left( \phi \right) =\max \{\left\vert \cos \left( \phi
+\alpha \right) \right\vert ,\alpha =0,2\pi /3,4\pi /3\},$ we get the
stochasticity condition
\begin{equation}
\left( 1+2\lambda _{2}\right) M\left( \phi \right) \leq 1\Leftrightarrow
0\leq \lambda _{2}\leq \frac{1}{2M\left( \phi \right) }-\frac{1}{2}
\end{equation}%
which gives a nonzero $\lambda _{2}$ for $M\left( \phi \right) <1,$ that is,
for$\ 0<\phi <\pi /3.$

So, we have to address the problem of stating a sufficient, and necessary,
condition for an assigned family of $n-$dimensional stochastic vectors $%
\left\{ \boldsymbol{\tau }\left( g_{j}\right) \right\} _{j=1}^{K}$ to be the
tomogram of a state with respect to a given $n-$dimensional irreducible
representation $D^{\alpha }$ of the group $G_{K}:$%
\begin{equation}
\exists \rho :\tau _{m}\left( g_{j}\right) =\left( V_{g_{j}}^{\alpha \dagger
}\rho V_{g_{j}}^{\alpha }\right) _{mm},\quad m=1,\dots ,n,\quad j=1,\dots ,K.
\end{equation}%
A sufficient (and necessary) condition can be stated in terms of positivity
of a suitable group function.

As the diagonal matrices $d_{g}^{\alpha }$'s depend only on representation $%
D^{\alpha }$ and are supposed to be known, we can define a normalized group
function $\psi ^{\alpha }$ as%
\begin{equation}
\psi ^{\alpha }(g_{j})=\sum\limits_{m=1}^{n}\mathrm{e}^{\mathrm{i}\theta
_{m}^{\alpha }(g_{j})}\tau _{m}(g_{j}).  \label{stocdistri}
\end{equation}

Requiring that the tomographic symbols $\left\{ v_{m}^{\alpha }\left(
g_{j}\right) \right\} $ of the operator
\begin{equation}
\frac{n_{\alpha }}{K}\sum\limits_{j=1}^{K}\sum\limits_{m=1}^{n}\mathrm{e}^{-%
\mathrm{i}\theta _{m}^{\alpha }(g_{j})}\tau _{m}(g_{j})D^{\alpha }(g_{j})
\label{ConstrOp}
\end{equation}%
constructed using this group function are just the assigned stochastic
vectors,\ the self-consistency relation $\left( \ref{selfconsist}\right) $
yields.%
\begin{equation}
\frac{n_{\alpha }}{K}\sum\limits_{j=1}^{K}\sum\limits_{m=1}^{n_{\alpha
}}R_{pm}^{\alpha }\left( g_{j},g_{h}\right) \tau _{m}\left( g_{j}\right)
=v_{p}^{\alpha }(g_{h})=\tau _{p}\left( g_{h}\right) .  \label{admissTau}
\end{equation}%
This is a necessary condition that the stochastic vectors must satisfy in
order to solve the posed problem, we may call it a condition of
compatibility of the $\tau $'s with the representation $D^{\alpha }$.

Besides, requiring that the operator $\left( \ref{ConstrOp}\right) $ is
self-adjoint gives%
\begin{equation}
\hskip -1.5cm \frac{n_{\alpha }}{K}\sum\limits_{j=1}^{K}\sum\limits_{m=1}^{n}%
\mathrm{e}^{-\mathrm{i}\theta _{m}^{\alpha }(g_{j})}\tau
_{m}(g_{j})D^{\alpha }(g_{j})=\frac{n_{\alpha }}{K}\sum\limits_{j=1}^{K}\sum%
\limits_{m=1}^{n}\mathrm{e}^{-\mathrm{i}\theta _{m}^{\alpha
}(g_{j}^{-1})}\tau _{m}(g_{j})D^{\alpha }(g_{j}^{-1}) \, .
\end{equation}

Finally, we check wether $\psi ^{\alpha }$ is a positive-type function. If
the answer to the check is in the affirmative, the observable $\left( \ref%
{ConstrOp}\right) $\ is just a density state $\rho _{\tau }^{\alpha }\,$
\begin{equation}
\rho _{\tau }^{\alpha }:=\frac{n}{K}\sum_{j=1}^{K}(\psi ^{\alpha
}(g_{j}))^{\ast }D^{\alpha }(g_{j}),  \label{rhotau}
\end{equation}%
such that its tomogram with respect to $D^{\alpha }$\ is just the assigned
family of stochastic vectors:
\begin{equation}
\left\{ \boldsymbol{W}^{\alpha }\left( g_{j}\right) \right\}
_{j=1}^{K}=\left\{ \boldsymbol{\tau }\left( g_{j}\right) \right\} _{j=1}^{K}.
\label{tomstoch}
\end{equation}%
In this case, we call tomogram the given family. Equivalently, we can write:
\begin{equation}
\psi ^{\alpha }(g_{j})=\mathrm{Tr}\left[ \rho _{\tau }^{\alpha }D^{\alpha
}\left( g\right) \right] .
\end{equation}%
So, the positivity condition implies that the stochastic family is
compatible with $D^{\alpha }$ and this, in turn, implies that, in the
decomposition \ of a group function with respect to the matrix elements of$\
$all the irreducible representations, the normalized function $\psi ^{\alpha
}$ has only components in the representation $D^{\alpha }.$ This completes
the proof.

\textbf{Example.} We now illustrate the above analysis with the example of
the $D^{2}$ representation of the triangle group.

The more general stochastic $2-$dimensional distribution on the group reads

\begin{equation}
\tau (g_{j})=\frac{1}{2}%
\left[\begin{array}{c}   
1+x_{j} \\
1-x_{j}%
\end{array}\right]%
,\quad -1\leq x_{j}\leq 1,\quad j=1,2,\dots ,6.
\end{equation}

The compatibilty condition with $D^{2}$ yields%
\begin{equation}
x_{4}+x_{5}+x_{6}=0.
\end{equation}

The Hermiticity condition gives%
\begin{equation}
x_{2}=x_{3}.
\end{equation}%
Construct the group function $\psi ^{2}$\ using eq.$\left( \ref{stocdistri}%
\right) $ and the above self- consistency and Hermiticity relations. The
Naimark matrix $\psi ^{2}(g_{i}^{-1}g_{j}),i,j=1,\dots ,6,$\ has the
following distinct eigenvalues:%
\begin{equation}
0,\quad \frac{3}{2}\pm \frac{1}{2}\sqrt{3\left(
3x_{2}^{2}+4x_{5}x_{6}+4x_{5}^{2}+4x_{6}^{2}\right) }.
\end{equation}%
Positivity requires that%
\begin{equation}
3x_{2}^{2}+4x_{5}x_{6}+4x_{5}^{2}+4x_{6}^{2}\leq 3
\end{equation}%
This constraint can be easily understood after diagonalization, putting:%
\begin{equation}
x_{5}+x_{6}=x,\quad x_{5}-x_{6}=\sqrt{3}y,\quad x_{2}=z,
\end{equation}%
which yields%
\begin{equation}
x^{2}+y^{2}+z^{2}=r^{2}\leq 1,
\end{equation}%
allowing the identification with the condition satisfied by density states
in two dimensions, discussed in subsec.7.2. In other words, there exists a
one-to-one correspondence between density states and stochastic
distributions satisfying positivity condition, which result just their
tomograms.

In conclusion, in the space of parameters $\left\{ -1\leq x_{j}\leq
1\right\} ,j=1,\dots ,6,$ the relations $\ \left\{ x_{2}=x_{3},\quad
x_{4}=-x_{5}-x_{6}\right\} $ define the set of stochastic vectors in one to
one correspondence with tomographic symbols of observables in the
representation $D^{2}$, which contains the unit ball of density states
defined by the constraint $3x_{2}^{2}+4x_{5}x_{6}+4x_{5}^{2}+4x_{6}^{2}\leq
3.$

Now, to conclude this example, choose a tomographic family of stochastic
vectors by means of a suitable point $\left( x,y,z\right) ,$ corresponding
to the density state%
\begin{equation}
\rho =\frac{1}{2}\left(
\begin{array}{cc}
1+z & x-\mathrm{i}y \\
x+\mathrm{i}y & 1-z%
\end{array}%
\right)
\end{equation}%
which is diagonalized, when $x+\mathrm{i}y\neq 0,$ by\ the unitary matrix $u$%
\begin{equation}
u=\frac{1}{\sqrt{2}}\left(
\begin{array}{cc}
\left( z-r\right) \left( r^{2}-rz\right) ^{-\frac{1}{2}} & \left( z+r\right)
\left( r^{2}+rz\right) ^{-\frac{1}{2}} \\
\left( x+\mathrm{i}y\right) \left( r^{2}-rz\right) ^{-\frac{1}{2}} & \left(
x+\mathrm{i}y\right) \left( r^{2}+rz\right) ^{-\frac{1}{2}}%
\end{array}%
\right)
\end{equation}%
The matrix $u$ corresponding to the diagonal case $x+\mathrm{i}y=0$ cannot
be obtained by a limit procedure.

In view of the Naimark theorem and construction of sec.6, it is possible to
exhibit explicit formulae for a unitary representation and a pure cyclic
vector state $\xi $ to represent canonically the $\psi ^{2}$ corresponding
to the chosen point $\left( x,y,z\right) $.

One gets a four dimensional Hilbert space, acted upon by the following
reducible representation of the group $S_{3}$
\begin{equation}
\left(
\begin{array}{cc}
D^{2} & 0 \\
0 & D^{2}%
\end{array}%
\right)
\end{equation}%
and the following density matrix for the pure cyclic state
\begin{equation}
\rho _{\xi }=U\left(
\begin{array}{cccc}
\rho _{-} & 0 & 0 & \sqrt{\rho _{-}\rho _{+}} \\
0 & 0 & 0 & 0 \\
0 & 0 & 0 & 0 \\
\sqrt{\rho _{-}\rho _{+}} & 0 & 0 & \rho _{+}%
\end{array}%
\right) U^{\dagger }.
\end{equation}%
Here $\rho _{\mp }=\frac{1}{2}\left( 1\mp r\right) $ are the eigenvalues of $%
\rho $ and $U$ is a $4\times 4-$ matrix in block-form
\begin{equation}
U=\left(
\begin{array}{cc}
u & 0 \\
0 & u%
\end{array}%
\right) .
\end{equation}%
One can check that the Hermitian matrix $\rho _{\xi }$ has trace one and $%
\mathrm{Tr}\left[ \rho _{\xi }^{2}\right] =1$ so it is the density matrix of
a pure state $\xi $. Since $U$ and $\rho $ are explicitly given in terms of
the stochastic distribution $\tau \left( x,y,z\right) ,$ we got the relation
between tomographic probability distributions on the group and Naimark pure
cyclic vector states $\xi .$

Back to the general finite group case, suppose there are two (or more)
irreducible different representation $D^{\alpha },D^{\beta }$ with the same
dimensionality $n_{\alpha _\beta },$ and that the stochastic vectors $%
\left\{ \boldsymbol{\tau }^{\alpha }(g_{j})\right\} _{j=1}^{K},$
corresponding to a state $\rho _{\tau }^{\alpha },$ make positive the
function:%
\begin{equation}
\psi ^{\alpha }(g_{j})=\sum\limits_{m=1}^{n}\mathrm{e}^{\mathrm{i}\theta
_{m}^{\alpha }(g_{j})}\tau _{m}^{\alpha }(g_{j}).
\end{equation}

We can construct the set of stochastic vectors $\left\{ \boldsymbol{\tau }%
^{\beta }(g_{j})\right\} _{j=1}^{K}$ corresponding to the same state $\rho
_{\tau }^{\alpha }\ $and making positive the function:%
\begin{equation}
\psi ^{\beta }(g_{j})=\sum\limits_{m=1}^{n}\mathrm{e}^{\mathrm{i}\theta
_{m}^{\beta }(g_{j})}\tau _{m}^{\beta }(g_{j})=\mathrm{Tr}\left[ \rho _{\tau
}^{\alpha }D^{\beta }\left( g\right) \right] .
\end{equation}%
In fact, in view of the reconstruction formula $\left( \ref{recform}\right)
, $ we have:%
\begin{eqnarray}
&&\frac{n_{\alpha }}{K}\sum\limits_{j=1}^{K}\sum\limits_{m=1}^{n_{\alpha }}%
\mathrm{e}^{-\mathrm{i}\theta _{m}^{\alpha }\left( g_{j}\right) }\tau
_{m}^{\alpha }(g_{j})\sum\limits_{r,s=1}^{n_{\alpha }}\left(
V_{g_{h}}^{\beta \dagger }\right) _{pr}D_{rs}^{\alpha }\left( g_{j}\right)
\left( V_{g_{h}}^{\beta }\right) _{sp} \\
&=&\sum\limits_{r,s=1}^{n_{\alpha }}\left( V_{g_{h}}^{\beta \dagger }\right)
_{pr}\rho _{rs}\left( V_{g_{h}}^{\beta }\right) _{sp}=\tau _{p}^{\beta
}(g_{h}).  \nonumber
\end{eqnarray}

Now, it can happen that the same family of stochastic vectors satisfies the
positivity condition of both the group functions $\psi ^{\alpha },\psi
^{\beta }.$ Then, in view of eq.(\ref{tomstoch}), the tomograms $\boldsymbol{%
W}^{\alpha },\boldsymbol{W}^{\beta }$\ are the same and two possibilities
can present: or $V^{\alpha }=V^{\beta }$ either $V^{\alpha }\neq V^{\beta }$
for any group element $g_{j}.$

In the first case, in view of eq.(\ref{rhotau}), the reconstructed density
states the same: $\rho _{\tau }^{\alpha }=\rho _{\tau }^{\beta }.$ For
example, this is the case of the two inequivalent $2-$dimensional
representations of the tetrahedron group, related to the $D^{2}$
representation of the triangle group $S_{3}$\ as $D^{\alpha
}=\{D^{2},D^{2}\} $ and $D^{\beta }=\{D^{2},-D^{2}\}$.

In the second case, the states are different: $\rho _{\tau }^{\alpha }\neq
\rho _{\tau }^{\beta }.$ This is the case, for instance, of the $3-$%
dimensional irreducible representations of $SU\left( 3\right) $, $D^{\alpha
}=D,D^{\beta }=D^{\ast },$ where $\rho _{\tau }^{\beta }=\rho _{\tau
}^{\alpha \ast }\neq \rho _{\tau }^{\alpha }.$ This result is obtained by a
straightforward and obvious generalization of all the above formulae and
conditions to the case of compact groups.

Briefly, given on the group $G$\ an irreducible representation $D$ and the
stochastic vector function $\left\{ \tau _{\left\{ m_{b};m_{c}\right\}
}\left( g\right) \right\},$ whose components are labelled by using a
suitable Gelfand-Zetlin basis, one can construct the group function
\begin{equation}
\psi \left( \tilde{g}\right) =\sum\limits_{\left\{ m_{b};m_{c}\right\} }\exp
(-\mathrm{i}\xi ^{b}m_{b})\tau _{\left\{ m_{b};m_{c}\right\} }\left(
g\right) .
\end{equation}%
By using eq. $\left( \ref{compRec}\right) ,$ a density state $\rho $\ can be
recovered by $\psi \left( \tilde{g}\right) $ \textsl{iff} this function is
of positive-type. Moreover, if the stochastic vector function is compatible
with $D,$ it is the tomogram of $\rho :$
\begin{equation}
{W}_{\rho }(g\,;\left\{ m_{b};m_{c}\right\} )=\tau _{\left\{
m_{b};m_{c}\right\} }\left( g\right) ,
\end{equation}%
and this solves completely the inverse tomographic problem.

Compatibility condition may be written as
\begin{eqnarray}
&&\tau _{\left\{ m_{b}^{\prime };m_{c}^{\prime }\right\} }\left( g^{\prime
}\right) =d^{\left( D\right) }\int\limits_{G}\psi \left( \tilde{g}\right)
^{\ast }\left( D^{\dagger }(g^{\prime })D(\tilde{g})D(g^{\prime })\right)
_{\left\{ m_{b}^{\prime };m_{c}^{\prime }\right\} \left\{ m_{b}^{\prime
};m_{c}^{\prime }\right\} }d\tilde{g} \\
&&=d^{\left( D\right) }\int\limits_{G}\sum\limits_{\left\{
m_{b};m_{c}\right\} }\mathrm{e}^{-\mathrm{i}\xi ^{b}m_{b}}\tau _{\left\{
m_{b};m_{c}\right\} }\left( g\right) \left( D^{\dagger }(g^{\prime })D(%
\tilde{g})D(g^{\prime })\right) _{\left\{ m_{b}^{\prime };m_{c}^{\prime
}\right\} \left\{ m_{b}^{\prime };m_{c}^{\prime }\right\} }d\tilde{g}  \nonumber
\end{eqnarray}%
where $\tilde{g}=g\exp (\xi ^{b}H_{b})g^{-1}$ and $H_{b}$'s are the
generator of the Cartan subalgebra, as usual.

We remark that checking the positivity of a compact group function\ like the
above $\psi \left( \tilde{g}\right) $\ amounts to an infinite number of
operations.

However, if an irreducible representation $D(G_{K})$ of a finite group can
be found in $D(G),$ then one can limit to check the positivity condition on
the finite group only for one $K\times K-$matrix. Assume that this holds
true. For example, this is the case of the defining representation of $%
U\left( 2\right) ,$ which contains the representation $D^{2}$ of the group $%
S_{3}.$

Besides, suppose that $\psi \left( \tilde{g}\right) $ satisfies the
compatibility condition with $D,$ so that it has no components with respect
to other irreducible representations. In this situation the positivity of $%
\psi $ on $G$ can be checked on $G_{K}$.

In fact, if $\psi $ is positive on $G_{K}$, we get a density state $\rho $
on the $n-$dimensional Hilbert space on which $D$ acts such that
\begin{equation}
\psi (g_{j})=\mathrm{Tr}[\rho D(g_{j})]=\sum\limits_{r,s=1}^{n}\rho
_{sr}D_{rs}(g_{j}).
\end{equation}%
By hypothesis $\psi $ can be expanded using only the matrix elements of $D$:%
\begin{equation}
\psi (g)=\sum\limits_{r,s=1}^{n}c_{rs}D_{rs}(g),
\end{equation}%
that are orthogonal on $G$ as well on $G_{K}$
\begin{equation}
\delta _{r,q}\delta _{s,p}=\frac{n}{K}\sum\limits_{j=1}^{K}D_{rs}^{\ast
}(g_{j})D_{qp}(g_{j})=d^{\left( D\right) }\int\limits_{G}D_{rs}^{\ast
}(g)D_{qp}(g)dg.
\end{equation}%
It readily follows that%
\begin{equation}
c_{rs}=\rho _{sr}\Rightarrow \psi (g)=\mathrm{Tr}[\rho D(g)]
\end{equation}%
and $\psi (g)$ is positive on $G.$

\section{Conclusions}

To conclude, we summarize the main results of our work. For states of finite
dimensional $C^{\star}-$algebras we have introduced the notion of
tomographic probability distribution. This concept provides the possibility
of clarifying new aspects of $C^{\star}-$algebras related to information
characteristics of the probability distributions\ like different kinds of
entropies.

These tomograms were also introduced for finite and compact groups by using
known unitary finite dimensional irreducible representations of these
groups. The tomographic probability vectors (tomograms) introduced for those
groups were shown to contain complete information on the quantum states
(Hermitian, trace-class, nonnegative matrices) associated with the
irreducible unitary representations of those groups.

The notion of Naimark matrix and its properties were used to study necessary
and sufficient conditions for the stochastic vectors defined on the finite
or compact groups to be tomographic probability distributions. The Naimark
theorem on positive-type group functions was shown to play a key role in the
problem of connecting the tomographic probability vectors on the group with
the density states on the Hilbert space of the irreducible representations
of the group.

The paradigmatic examples of two groups, the group $S_3$ of permutations of
three points and $SU( 3 ) ,$ were discussed in detail. The general
construction of the $U\left( n\right) $ group (and other classical groups)
tomograms was presented by using the Gelfand-Zetlin basis labels of the
tomographic probability vectors.

The notion of tomographic probabilities introduced for finite $C^{\star }-$%
algebras was, in fact, shown to coincide with that of tomographic
probability vectors associated to finite unitary groups. The probability
vectors defined on finite or compact groups establish a relation between the
group structure and the structure of the simplexes containing those
probability vectors. An analogous relation exists between finite $C^{\star
}- $algebras and those simplexes, thanks to the existence of tomographic
probability vectors defined on the $C^{\star }-$algebras.

Thus, for finite and compact groups, their group algebras and abstract $%
C^{\star }-$algebras were considered in the unifying framework of the
tomographic approach, where the tomograms provide the possibility to
describe completely all the kinds of quantum states, both pure and mixed
ones.

For example, the spin states (qu-dits) associated with $SU\left( 2\right) -$%
group irreducible representations can be alternatively described by the
spin-tomographic probability distributions of measurable spin-projections on
the quantization axes.

We will develop these aspects of the tomographic approach to systems with
the discussed finite or compact symmetry groups in future papers.

\ack A. Ibort would like to acknowledge the partial support from MICIN
research project MTM2010-21186-C02-02 and QUITEMAD project P2009/ESP-1594.
V. I. Man'ko would like to thank I. N. F. N. and University "Federico II" of
Naples for their hospitality and R. F. F. I. for partial support. G. Marmo
would like to thank the support provided by the Banco de Santander-UCIIIM
\textquotedblleft Chairs of Excellence\textquotedblright\ Programme
2010-2011.  

\end{document}